\documentclass[prd,superscriptaddress,amsfonts,amssymb,amsmath,showpacs,twocolumn]{revtex4-1}
\usepackage{bm}
\usepackage{amsfonts}
\usepackage{latexsym}
\usepackage[latin1]{inputenc}
\usepackage{graphicx}
\usepackage{amsmath}
\usepackage{palatino}
\usepackage{mathpazo}
\usepackage{textcomp}
\usepackage{booktabs}
\usepackage{dcolumn}
\usepackage{hyperref}
\hypersetup{colorlinks,citecolor=blue}
\usepackage{amsmath}
\usepackage{xcolor}

\def\jnl@style{\it}
\def\aaref@jnl#1{{\jnl@style#1}}

\def\aaref@jnl#1{{\jnl@style#1}}

\def\aj{\aaref@jnl{AJ}}                   
\def\apj{\aaref@jnl{ApJ}}                 
\def\apjl{\aaref@jnl{ApJ}}                
\def\apjs{\aaref@jnl{ApJS}}               
\def\apss{\aaref@jnl{Ap\&SS}}             
\def\aap{\aaref@jnl{A\&A}}                
\def\aapr{\aaref@jnl{A\&A~Rev.}}          
\def\aaps{\aaref@jnl{A\&AS}}              
\def\mnras{\aaref@jnl{Mon.~Not.~Roy.~Astron.~Soc.}}             
\def\prd{\aaref@jnl{Phys.~Rev.~D}}        
\def\prc{\aaref@jnl{Phys.~Rev.~C}}  
\def\prl{\aaref@jnl{Phys.~Rev.~Lett.}}    
\def\qjras{\aaref@jnl{QJRAS}}             
\def\skytel{\aaref@jnl{S\&T}}             
\def\ssr{\aaref@jnl{Space~Sci.~Rev.}}     
\def\zap{\aaref@jnl{ZAp}}                 
\def\nat{\aaref@jnl{Nature}}              
\def\aplett{\aaref@jnl{Astrophys.~Lett.}} 
\def\apspr{\aaref@jnl{Astrophys.~Space~Phys.~Res.}} 
\def\physrep{\aaref@jnl{Phys.~Rep.}}      
\def\physscr{\aaref@jnl{Phys.~Scr}}       
\def\commat{\aaref@jnl{Comm.~Math.~Phys.}}              
\def\science{\aaref@jnl{Science}}               
\def\cqg{\aaref@jnl{Classical Quant.~Grav.}}            
\def\jpcs{\aaref@jnl{JPCS}}                                     
\def\ijmpd{\aaref@jnl{Int.~J.~Mod.~Phys.~D}}                    
\def\grg{\aaref@jnl{Gen.~Relat.~Gravit.}}               
\def\rpp{\aaref@jnl{Rep.~Prog.~Phys.}}          
\def\npa{\aaref@jnl{Nucl.~Phys.~A}}        
\def\lrr{\aaref@jnl{Living Rev.~Rel.}}                   
\def\jcap{\aaref@jnl{J.~Cosmology Astropart.~Phys.}}    
\def\rmp{\aaref@jnl{Rev.~Mod.~Phys.}}   

\allowdisplaybreaks[1]

\begin{document}
\color{red}

\title{Wormholes in 4D Einstein-Gauss-Bonnet Gravity}
\author{Kimet Jusufi}
\email{kimet.jusufi@unite.edu.mk}
\affiliation{Physics Department, State University of Tetovo,  Ilinden Street nn, 1200, Tetovo, North Macedonia}
\affiliation{Institute of Physics, 
  Faculty of Natural Sciences and Mathematics,
  Ss. Cyril and Methodius University,
  Arhimedova 3, 1000 Skopje, North Macedonia}

\author{Ayan Banerjee}
\email{ayan\_7575@yahoo.co.in}
\affiliation{Astrophysics and Cosmology Research Unit, University of KwaZulu Natal, Private Bag X54001, Durban 4000,
South Africa.}

 \author{Sushant G. Ghosh}
 \email{sgghosh@gmail.com, sghosh2@jmi.ac.in}
\affiliation {Centre for Theoretical Physics, Jamia Millia Islamia, New Delhi 110025, India}
\affiliation{Astrophysics and Cosmology Research Unit, University of KwaZulu Natal, Private Bag X54001, Durban 4000,
South Africa.}
\begin{abstract}
Recent times witnessed a significant interest in regularizing, a $ D \to 4 $ limit, of EGB gravity initiated by  Glavan and Lin [Phys. Rev. Lett. 124, 081301 (2020)] by re-scaling GB coupling constant as $\alpha/(D-4)$ and taking limit $D \to 4$, and in turn these regularized $4D$ gravities have nontrivial gravitational dynamics.  Interestingly, the maximally or spherically symmetric solution to all the regularized  gravities coincides in the $4D$ case. In view of this, we obtain an exact spherically symmetric wormhole solution in the   $ 4D $ EGB gravity for an isotropic and anisotropic matter sources.  In this regard, we consider also a wormhole with a specific radial-dependent shape function, a power-law density profile as well as by imposing a particular equation of state. To this end, we analyze the flare-out conditions, embedding diagrams, energy conditions and the volume integral quantifier. In particular our $-$ve branch results, in the limit $\alpha \rightarrow 0$, reduced exactly  to  \emph{vis-$\grave{a}$-vis} 4D 
Morris -Thorne wormholes of GR.
 
\end{abstract}

\maketitle

\section{Introduction}\label{sec1}

Within the context of general relativity, wormholes are topological bridges connecting two different asymptotically at regions of spacetime \cite{Morris:1988cz,Visser} as well as two different asympotically de Sitter (dS) or anti-de Sitter (AdS) regions \cite{Lemos:2003jb}.  Interest in wormhole space-times dates back to 1916, when  Flamm \cite{flam} propose `tunnel structure' in the Schwarzschild solution represents a wormhole.  Paging through history one finds, Einstein and Rosen \cite{EinsteinRosen} proposed a ``bridge structure" that connect two exterior regions of a Schwarzschild black hole space time, and thus forms an inter-universe connection. This was the first attempts to obtain a regular solution without a singularity, namely, `Einstein-Rosen bridge' (ERB). However, it became soon clear that obtained solution was invalid particle model as mass-energy of such a  curved-space topology is the order of Planck mass. The term \textit{wormhole} for these bridges was first used by J. A. Wheeler \cite{Fuller1957,Fuller1962} for microscopic charge-carrying wormholes. They showed that 
 wormholes would collapse instantly upon formation. Moreover, if such a wormhole somehow opened, it would
 pinch off so quickly even a single photon could be transmitted through it, thereby preserving Einsteinian causality. Despite the dubious possibility of existing a wormhole solution, their study has opened up remarkably fruitful avenues of research. 

Modern interest has been focused on traversable Lorentzian wormholes (which have no horizons, allowing two-way passage through them),  was suggested by Morris and Thorne \cite{Morris:1988cz} and  subsequently Morris, Thorne and Yurtsever \cite{Morris:1988tu}. They started with static, spherically symmetric metric connecting two asymptotically flat spacetimes where matter and radiation can travel freely through it, and  is now a well known solution in general relativity. However, wormhole solutions are asymptotically flat with a constant or variable radius which  depends on its configuration.  Consequently,
these geometries have a minimal surface area linked to satisfy flare-out condition, called \emph{throat} of the wormhole.
For this property to be accomplished it is considered that the space-times require a stress-energy tensor that violates
 the weak/null energy conditions.  In classical GR this means, thatthe matter creating the wormhole must possess very exotic properties (negative-energy matter) \cite{Visser}, for example ghost scalar fields or phantom energy \cite{ArmendarizPicon:2002km,Sushkov:2005kj,Lobo:2005us}. These
hypothetical matter sounds to be unusual for the first time, but such matter appears in quantum field theory which appears as a natural consequence if the topology of spacetime fluctuates in time \cite{Wheeler}.

Since it is important and useful to minimize the usage of exotic matter. As such wormholes could be possible 
by choosing the geometry in a very specific and appropriate way, which was first pointed out by Visser \textit{et al.} \cite{Visser:2003yf} and further study have also reported \cite{Kuhfittig:2002ur}.
On the other hand, evolving wormhole satisfying the weak energy condition (WEC) could exist within classical GR \cite{Kar:1994tz,Kar:1995ss}. Even in GR it appears possible to avoids NEC violation in wormhole construction due to rotation in cylindrical symmetry, though separate effort is needed to achieve asymptotic flatness on both sides of the throat, see \cite{Bronnikov:2019hbl,Bronnikov:2019gsr}.

Subsequently, motivated by Morris Thorne \cite{Morris:1988cz} idea, there has been intense activities in the investigation of wormholes in the modified theories of gravity and also in the higher dimensional theories of gravity \cite{Harko:2013yb,Zangeneh:2014noa,Banerjee:2020uyi,Shaikh:2016dpl,Bronnikov:2015pha,Mehdizadeh:2018smu} including in the Kaluza-Klein gravity \cite{deLeon:2009pu,Dzhunushaliev:1998ya,Dzhunushaliev:2013iia}. Advantages in such theories are one can avoid nonstandard fluids and this was the main motivation for extensive research in modified gravity theories. Moreover, specific modifications of Einstein gravity allow additional degrees of freedom in the gravitational sector which can be used, amongst others, to resolve the dark energy and the dark matter problems. Wormholes in $f(R)$ gravity \cite{Lobo:2009ip} have been constructed without exotic matter. In the same context, Refs. \cite{Bah1,Mazharimousavi:2012xv,Pavlovic:2014gba,Sharif:2018jdj,DeBenedictis:2012qz} and the citations therein, are quite useful to understand wormhole geometries with different inputs and examined the validity of energy conditions. Such wormhole geometries are supported by a non-minimal curvature-matter coupling are obtained in \cite{MontelongoGarcia:2010xd} and also the curvature-matter coupled theory, $f(R,T)$ gravity, where exact wormholes solutions were obtained 
\cite{Moraes:2017mir,Elizalde:2018frj,Banerjee:2019wjj}. In Ref. \cite{Shaikh:2018yku}, wormhole solutions in the background of Born-Infeld theory, scalar-tensor tele-parallel theories \cite{Bah2} and other related works have been found \cite{Jusufi:2018waj,Jusufi:2019knb,Richarte:2007zz}.

The wormholes geometries also received significant attention in higher curvature Einstein-Gauss-Bonnet (EGB) theory \cite{Mehdizadeh:2015jra,Kanti:2011yv,Maeda:2008nz}, and also in the  Lovelock gravity \cite{Dehghani:2009zza,Zangeneh:2015jda}. 
It may be mentioned that EGB theory, particular case of  Lovelock gravity, is a natural generalizations of general relativity, to higher dimensions, introduced by Lanczos \cite{Lanczos:1938sf}, and rediscovered by David Lovelock \cite{Lovelock:1971yv,Lovelock:1972vz}. The EGB gravity has been widely studied, because it can be obtained in the low energy limit of string theory  \cite{Zwiebach:1985uq,Garraffo:2008hu}, is known free from instabilities when expanding about flat spacetime \cite{Boulware:1985wk}, and also leads to the ghost-free nontrivial gravitational self-interactions \cite{Nojiri:2018ouv}.  

However, the EGB theory is topological in $4D$ as the $GB$ Lagrangian is a total derivative, so it does not contribute to the gravitational dynamics and thereby for non-trivial gravitational dynamics in EGB theory one requires $D \geq 5$. 
This issue of the EGB theory was recently addressed by Glavan \& Lin \cite{Glavan:2019inb} by rescaling the Gauss-Bonnet coupling constant $\alpha$ as $\alpha/(D -4)$, and taking the limit $D \to 4$ at the level of the field equation and the resulting EGB in gives rise to non-trivial dynamics in $4D$. For definiteness we shall call it the   4D EGB gravity, which has some in interesting property viz. it bypasses the conclusions of Lovelock's theorem and avoids Ostrogradsky instability.
It worth pointing a priori that  dimensional regularization of this was considered by 
Tomozawa \cite{Tomozawa:2011gp} with similar consequences.

The 4D EGB theory received compelling attention initiated by Glavan and Lin \cite{Glavan:2019inb} (see also \cite{Cognola:2013fva,Tomozawa:2011gp}) who also proposed a static spherically symmetric vacuum black holes which hold interesting properties, e.g., the black holes are free from the singularity problem, at small distances, the gravitational force is repulsive and an infalling particle fails to reach the singularity \cite{Glavan:2019inb}. This is in contrast to the analogous $HD$ black holes \cite{Boulware:1985wk} Schwarzschild like curvature singularity inevitably forms.  Other cascades of work includes, Charged version of spherically
symmetric black holes \cite{Fernandes:2020rpa} in an anti-de Sitter spacetime \cite{Fernandes:2020rpa}, a Vaidya-like radiating black holes in Ref.~\cite{Ghosh:2020vpc}, generating black holes solution was also addressed in Ref. \cite{Ghosh:2020syx} also regular black holes \cite{Kumar:2020xvu,Kumar:2020uyz}.
Additionally, black hole solutions and their physical properties, such as rotating black holes using Newman-Janis algorithm \cite{Kumar:2020owy,Wei:2020ght}, rotating black hole as particle accelerator \cite{NaveenaKumara:2020rmi}. On the other hand, thermodynamical properties of anti-de Sitter black hole \cite{HosseiniMansoori:2020yfj},  geodesics motion and shadow \cite{Konoplya:2020bxa}, gravitational lensing \cite{Islam:2020xmy,Jin:2020emq}, relativistic stars in 4D EGB \cite{Doneva:2020ped}, and we refer the reader to \cite{Li:2020tlo,Mishra:2020gce,1,2,3,4,5,6,7,8,9} for other contributions and related issues details.  Important  contributions in the context include objections on the 4D EGB theory raised in Ref. \cite{Gurses:2020ofy} and the derivation of regularized field equations in Refs. \cite{Fernandes:2020nbq,Hennigar:2020lsl}.

However, the above regularization procedure proposed  in \cite{Glavan:2019inb,Cognola:2013fva}  is currently under debate as several question are also being raised regarding it's validity \cite{Ai:2020peo,Hennigar:2020lsl,Shu:2020cjw,Gurses:2020ofy,Mahapatra:2020rds,Arrechea:2020evj}.  In turn, alternate regularization procedures have been also proposed  \cite{Lu:2020iav,Kobayashi:2020wqy,Hennigar:2020lsl,Casalino:2020kbt, Kobayashi:2020wqy}.   However, the spherically symmetric $4D$  black hole  
solution obtained in \cite{Glavan:2019inb,Cognola:2013fva} remains are also coincides in these regularised theories \cite{Lu:2020iav,Hennigar:2020lsl,Casalino:2020kbt,Ma:2020ufk}, and no new solutions could be obtained in these alternate proposal at least for the special case of $4D$ Spherical symmetric spacestime.

However, the wormhole geometries is still unexplored, e.g., the generalization of Moris-Thorne wormhole solutions \cite{Morris:1988cz} is still unknown. It is the purpose of this paper to obtain   this wormhole in the    $4D$ EGB theory of gravity and investigate how the squared curvature terms effect the wormhole geometriy.  

 The paper is organized as follows:  In the next Sec. \ref{sec2}, we review the  field equations in the   4D EGB gravity and show that it makes a nontrivial contribution to gravitational dynamics in 4D as well.  In Sec. \ref{sec3}, among other things,  we obtain the field equations and four wormhole solutions including ans exact isotropic/anisotropic  wormhole spacetime in the   4D EGB gravity with
a constant redshift function. In Sec. \ref{mf} we discuss the wormhole mass function. The embedding diagrams of our wormhole metrics is the subject of Sec. \ref{sec4}. The Sec. \ref{sec5} is devoted to elaborate the energy conditions and the volume integral quantifier in \ref{sec6}. Finally we comment on our results in Sec. \ref{sec7}.
 
We have used units which fix the speed of light and the gravitational constant via $\hbar=G=c=1$.\\
\section{Basic equations of EGB gravity}\label{sec2}
We begin with a short review on the  EGB gravity in $D$-dimensions and also derive the equations of motion. The gravitational action of the EGB theory reads
\begin{equation}\label{action}
	\mathcal{I}_{A}=\frac{1}{16 \pi}\int d^{D}x\sqrt{-g}\left[ R +\frac{\alpha}{D-4} \mathcal{L}_{\text{GB}} \right]
+\mathcal{S}_{\text{matter}},
\end{equation}
where $g$ denotes the determinant of the metric $g_{\mu\nu}$ and $\alpha$ is the Gauss-Bonnet coupling coefficient with
dimension $[length]^2$. The discussion, in this paper,  will be given here corresponding to the case with $\alpha \ge 0$. The term $\mathcal{L}_{\text{GB}}$ is the Lagrangian defined by
\begin{equation}
	\mathcal{L}_{\text{GB}}=R^{\mu\nu\rho\sigma} R_{\mu\nu\rho\sigma}- 4 R^{\mu\nu}R_{\mu\nu}+ R^2\label{GB}.
\end{equation}
Here, $S_{\text{matter}}$ is the matter fields appearing in the theory. The variation of (\ref{action}) with respect to metric $g_{\mu \nu}$ gives the field equations \cite{Ghosh:2020vpc}
\begin{equation}\label{GBeq}
	G_{\mu\nu}+\frac{\alpha}{D-4} H_{\mu\nu}= 8 \pi T_{\mu\nu},
\end{equation}
where  $ T_{\mu\nu}= -\frac{2}{\sqrt{-g}}\frac{\delta\left(\sqrt{-g}\mathcal{S}_m\right)}{\delta g^{\mu\nu}}$ is the energy momentum tensor of matter with the following expression
\begin{eqnarray}
	G_{\mu\nu}&=&R_{\mu\nu}-\frac{1}{2}R g_{\mu\nu},\nonumber\\\notag
	H_{\mu\nu}&=&2\Bigr( R R_{\mu\nu}-2R_{\mu\sigma} {R}{^\sigma}_{\nu} -2 R_{\mu\sigma\nu\rho}{R}^{\sigma\rho} - R_{\mu\sigma\rho\delta}{R}^{\sigma\rho\delta}{_\nu}\Bigl)\\
	&-& \frac{1}{2}\mathcal{L}_{\text{GB}}g_{\mu\nu},\label{FieldEq}
\end{eqnarray}
with $R$ the Ricci scalar, $R_{\mu\nu}$ the Ricci tensor, $H_{\mu\nu}$ is the Lancoz tensor and $R_{\mu\sigma\nu\rho}$ the Riemann tensor.  In general  the GB terms is total derivative  in 4D space-time, and hence  do not contribute to the field equations. 
However, with re-scaled  coupling constant $ \alpha/(D-4)$, and  considering maximally symmetric spacetimes with curvature scale ${\cal K}$ \cite{Ghosh:2020vpc}, we obtain
\begin{equation}\label{gbc}
\frac{g_{\mu\sigma}}{\sqrt{-g}} \frac{\delta \mathcal{L}_{\text{GB}}}{\delta g_{\nu\sigma}} = \frac{\alpha (D-2) (D-3)}{2(D-1)} {\cal K}^2 \delta_{\mu}^{\nu},
\end{equation}
obviously the variation of the GB action does not vanish in $D=4$ because of the re-scaled coupling constant \cite{Glavan:2019inb}. \\

To obtain wormholes in the $ 4D $ EGB, we use the regularization process in \cite{Glavan:2019inb,Cognola:2013fva}, as the 4D spherical solutions obtained in   \cite{Glavan:2019inb,Cognola:2013fva} are also exactly same as obtained in other regularised theories \cite{Lu:2020iav,Hennigar:2020lsl,Casalino:2020kbt,Ma:2020ufk}.

\section{Wormhole solutions for the   4$D$ EGB}\label{sec3}

To begin discussion on the wormhole in the   4$D$ EGB,  it is mandatory to consider the general static, spherically symmetric metric $D$-dimensional metric \cite{Mehdizadeh:2015jra} given by 
\begin{equation}
ds^2=-e^{2\Phi(r)}dt^2+\frac{dr^2}{1-\frac{b(r)}{r}}+r^2d\Omega_{D-2}^2.\label{metric}
\end{equation} 
where
\begin{equation}
d\Omega^2_{D-2} = d\theta^2_1 + \sum^{D-2}_{i=2}\prod^{i-1}_{j=1}
\sin^{2}\theta_j\;d\theta^2_i \;,\notag
\end{equation}
and $\Phi(r)$ is the redshift functions of an infalling body, and it must be finite everywhere to avoid the
presence an event horizon. On the other hand, $b(r)$ represents the spatial shape function of the wormhole geometry, it determine the shape of the wormhole in the embedding diagram \cite{Mehdizadeh:2015jra}. 
Note that $b(r)$ should obey the boundary condition $b(r = r_0$) = $r_0$ at the throat $r_0$ where $r_0 \leq r \leq \infty$. 
Now, to ensure the traversibility of wormhole, the function $b(r)$ must satisfy the \textit{flaring-out} condition that can be obtained from the embedding calculation, and reads
\begin{equation}\label{a12}
\frac{b(r)-rb^{\prime}(r)}{b^2(r)}>0.
\end{equation}\\

This condition can also be  written in a compactified form, namely,
$b^{\prime}(r_0) < 1$ at the throat $r = r_0$. The condition $1 - b(r)/r \geq 0$ is also imposed.

We consider an anisotropic fluid for the matter source defined by the stress energy tensor
\begin{eqnarray}
T^\nu_i &=& (\rho+\mathcal{P}_t)u^\nu u_i + \mathcal{P}_t g^\nu_i + (\mathcal{P}_r-\mathcal{P}_t)\chi_i \chi^\nu, \label{eq8}
\end{eqnarray}
where $u_\nu$ is the $D$-velocity and $\chi_\nu$ is the unit spacelike vector in the radial direction with $\rho(r)$ is the energy density and $\mathcal{P}_r (r)$ and $\mathcal{P}_t (r)$  are the radial and transverse pressures, respectively. On using the 
metric (\ref{metric}) with stress tensor (\ref{eq8}), in the limit $D \to 4$, the components of the field equations (\ref{GBeq}) can be written as
\begin{widetext}
\begin{eqnarray}\label{DRE1}
&& 8\pi \rho(r)= \frac{\alpha b(r)}{r^6}\left(2 r b'(r)-3b(r)\right)+\frac{b'(r)}{r^2} ,\\
&& 8\pi \mathcal{P}_r (r)= \frac{\alpha b(r)}{r^6}\left(4 \Phi' r(r-b(r))+b(r)\right)+\frac{2\Phi '(r-b(r))}{r^2}-\frac{b(r)}{r^3}, \label{DRE2}\\
&& 8\pi \mathcal{P}_t (r) = \left(1-\frac{b(r)}{r}\right)\left[\left(\Phi''+\Phi'^2\right)\left(1+\frac{4\alpha b(r)}{r^3}\right) 
+\frac{1}{r} 
\left(\Phi'-\frac{rb'(r)-b(r)}{2r(r-b(r))}\right)\left(1-\frac{2\alpha b(r)}{r^3}\right)  \nonumber \right. \\ 
  &&  \hspace{1.3 cm}  \quad \left.
- \frac{(rb'(r)-b(r)) \Phi'}{2r(r-b(r))} \left(1-\frac{8\alpha }{r^2}+\frac{12\alpha b(r)}{r^3}\right)\right]-\frac{2\alpha b^2(r)}{r^6},   \notag\\ \label{DRE3}
\end{eqnarray}
\end{widetext}
where the prime denotes a derivative with respect to the
radial coordinate $r$. In this context, we have five unknown functions of $r$, i.e., $\rho(r)$, $\mathcal{P}_r$, $\mathcal{P}_t$ (r), $b(r)$ and $\Phi(r)$. We provide below several plan of action for solving the system of equations. Further, we need an additional restriction to close the system and
solve the field equations.

\subsection{Isotropic solution}
In order to simplify the problem we consider a constant redshift function, namely, a wormhole solution with zero tidal force, i.e. $\Phi(r)= \Phi_0=\text{const}$, which simplifies the calculations, and provides interesting exact wormhole solutions. In a recent works \cite{Fernandes:2020nbq,Hennigar:2020lsl}, it was shown that by taking the trace of the field equations (3) one has the simple form
\begin{equation}\label{n12}
    R+\frac{\alpha}{2}\mathcal{L}_{\text{GB}}=-8\pi T,
\end{equation}
where the trace $T=T^{\mu}_{ \nu}$. Using the metric form (\ref{metric}) for isotropic fluid matter we can use the relation $
    \mathcal{P}_r(r)=\mathcal{P}_t(r) =\omega \rho(r)
$  obtain the following condition 
\begin{equation}
    \frac{2 b'(r)}{r^2}+8\pi \left[-\rho(r)+3 \omega \rho(r) \right]=0.
\end{equation}
Utilizing Eq. (\ref{n12}) and  (\ref{DRE1}), we obtain
a differential equation for the shape function as
\begin{eqnarray}\notag
    && 3 r \left[2 (\omega-\frac{1}{3})\alpha b(r)+r^3(\omega+\frac{1}{3})  \right] b'(r)\\
    &-& 9 b^2(r)\alpha (\omega-\frac{1}{3})=0.
\end{eqnarray}
Solving the above equation for $b(r)$
which has the following form
\begin{equation}
    b(r)=-\frac{r^3(3\omega+1)}{\alpha (3 \omega-1)}\left( 1 \pm \sqrt{1+\frac{4 C \alpha (3\omega-1)}{ r^3 (1+3\omega)^2}}    \right),
\end{equation}
and apply the condition $b(r=r_0)=r_0$, we find the constant of integration
\begin{equation}
    C=\frac{r_0^2(3\omega+1)+\alpha(3 \omega-1)}{r_0}
\end{equation}
Finally, the explicit form of shape function is
\begin{equation}\label{n17}
    b(r)=-\frac{r^3(3\omega+1)}{\alpha (3 \omega-1)}\left( 1 \pm \sqrt{1+\frac{4 \alpha \mathcal{A} }{ r^3 r_0 }  }    \right),
\end{equation}
where 
\begin{equation}
    \mathcal{A}=\frac{(3\omega-1)}{(1+3\omega)^2}( r_0^2(3 \omega+1)+\alpha(3 \omega-1)).
\end{equation}
Note that the solution given by (\ref{n17}) holds for $\omega  \neq 1/3$ only. The $\pm$ sign in Eq. (\ref{n17}) refers to two different branches of
solution. Boulware and Deser \cite{Boulware:1985wk} have demonstrated that EGB black
holes with $+$ve branch sign are unstable and the graviton degree of freedom is a ghost, while the branch with $-$ve sign is stable and is free of ghosts. 
In our case, in the limit $\alpha \to 0$, the  $+$ve positive branch leads to 
\begin{equation}
    \frac{b(r)}{r}=-\frac{r^2(3 \omega+1)}{(3\omega-1)\alpha}-\frac{r_0}{r} \ldots
\end{equation}
which is a wormhole solution in a de-Sitter/ anti-de Sitter spacetimes depending on the sign of $\alpha$. On the other hand, in the limit $\alpha \to 0$, the  $-$ve  goes over
\begin{equation}
    \frac{b(r)}{r}=\frac{r_0}{r}+ \ldots,
\end{equation}
 and the standard Morris-Thorne wormhole is obtained when the cosmological constant vanishes. Henceforth we restrict to the $-$ve branch in which case the 4D EGB wormhole metric reads 
\begin{equation}\label{metricw}
    ds^2=-dt^2+\frac{dr^2}{1+\frac{r^2(3\omega+1)}{\alpha (3 \omega-1)}\left( 1 \pm \sqrt{1+\frac{4 \alpha \mathcal{A} }{ r^3 r_0 }  }    \right) }  +r^2 d\Omega^2_2,
\end{equation}
with the unit sphere line element is given by  $d \Omega_2^2$= $d \theta^2+\sin^2\theta d \phi^2$. Notice that the above analyses is correct for the interval $\omega \geq -1/3$ along with  $\omega  \neq 1/3$.  For values $\omega <-1/3$ the sign of the solutions flips.

\subsection{Supporting conditions in anisotropic fluid scenario}
We shall argue that there are two ways to find an anisotropic solution. The first way is to use Eq. (\ref{n12}), while the second way is to utilize the conservation of the energy-momentum tensor. As one can easily see that for  static and spherical symmetric equations  of motion with the anisotropy fluid is in hydrostatic equilibrium. Therefore, 
covariant derivative of the energy momentum tensor of  matter is given by 
\begin{equation}\label{aniso}
  \mathcal{P'}_r(r)=\frac{2(\mathcal{P}_t(r)-\mathcal{P}_r(r))}{r}-(\rho(r)+\mathcal{P}_r(r))\Phi'(r).
\end{equation}

 Taking into account Eqs. (\ref{DRE2}-\ref{DRE3}) along with an interesting EoS \cite{moraes},
\begin{equation}\label{n23}
\mathcal{P}_t(r)=\omega_t \mathcal{P}_r(r),
\end{equation}
  leads to the following expression
\begin{eqnarray}\notag
    \left( 2 r \alpha b(r)-r^4  \right)b'(r)&+& 2 b(r) r^3 (\omega_t+\frac{1}{2})\\
    &-& 2 b(r)\alpha b(r)(\omega_t+2)=0,
\end{eqnarray}
where the EoS parameter $\omega_t$ and the redshift function $\Phi(r)$ are constants. Solving the last differential equation we get the following shape function 
\begin{equation}\label{n25}
    b(r)=\frac{r^3}{2 \alpha}\left[1 \pm \sqrt{1+ \frac{4 \alpha e^{-2 C_1 (\omega_t-1)}}{r^{2(1-\omega_t)}}}    \right],
\end{equation}
where $C_1$ is a constant of integration, which can be fixed using the condition $b(r=r_0)=r_0$, Then the shape function modifies to
\begin{equation}\label{n26}
    b(r)=\frac{r^3}{2 \alpha} \left[1 \pm \sqrt{1+\frac{4 \alpha r^{2(\omega_t-1)} }{r_0^{2 (\omega_t+1)}}(\alpha-r_0^2)}  \right].
\end{equation}

Similarly for the $+$ve branch sign in the limit $\alpha \to 0$, we obtain
\begin{equation}
    \frac{b(r)}{r}=-\left(\frac{r_0}{r}\right)^{-2 \omega}+ \frac{r^2}{\alpha }+...
\end{equation}
which is a wormhole solution in a de-Sitter/ anti-de Sitter spacetimes depending on the sign of $\alpha$. On the other hand, in the limit $\alpha \to 0$, the  $-$ve  goes over
\begin{equation}
    \frac{b(r)}{r}=\left(\frac{r_0}{r}\right)^{-2 \omega}+ \ldots,
\end{equation}
 which reduces to the standard Morris-Thorne wormhole. Henceforth, we restrict to the $-$ve branch for anisotropic fluid in the   4D EGB  takes the form
\begin{equation}\label{eq39}
    ds^2=-dt^2+\frac{dr^2}{1-\frac{r^2}{2 \alpha} \left[1 - \sqrt{1+\frac{4 \alpha r^{2(\omega_t-1)} }{r_0^{2 (\omega_t+1)}}(\alpha-r_0^2)}   \right]}+r^2d\Omega_2^2.
\end{equation}

As we have already pointed out, the second way to obtain our solution (\ref{n26}) is based on the condition (\ref{n12}). Thus, using  equations (\ref{n23}) and (\ref{n12}), we will re-derive the expression
\begin{equation}
    \frac{2 b'(r)}{r^2}+8\pi \left[-\rho(r)+\mathcal{P}_r(1+ 2\omega_t) \right]=0.
\end{equation}
Substituting all the expressions (\ref{DRE1}-\ref{DRE3})
and solve for $b(r)$, one can independently arrive the 
same solution (\ref{n25}).

\begin{figure*}
\includegraphics[width=8.0 cm]{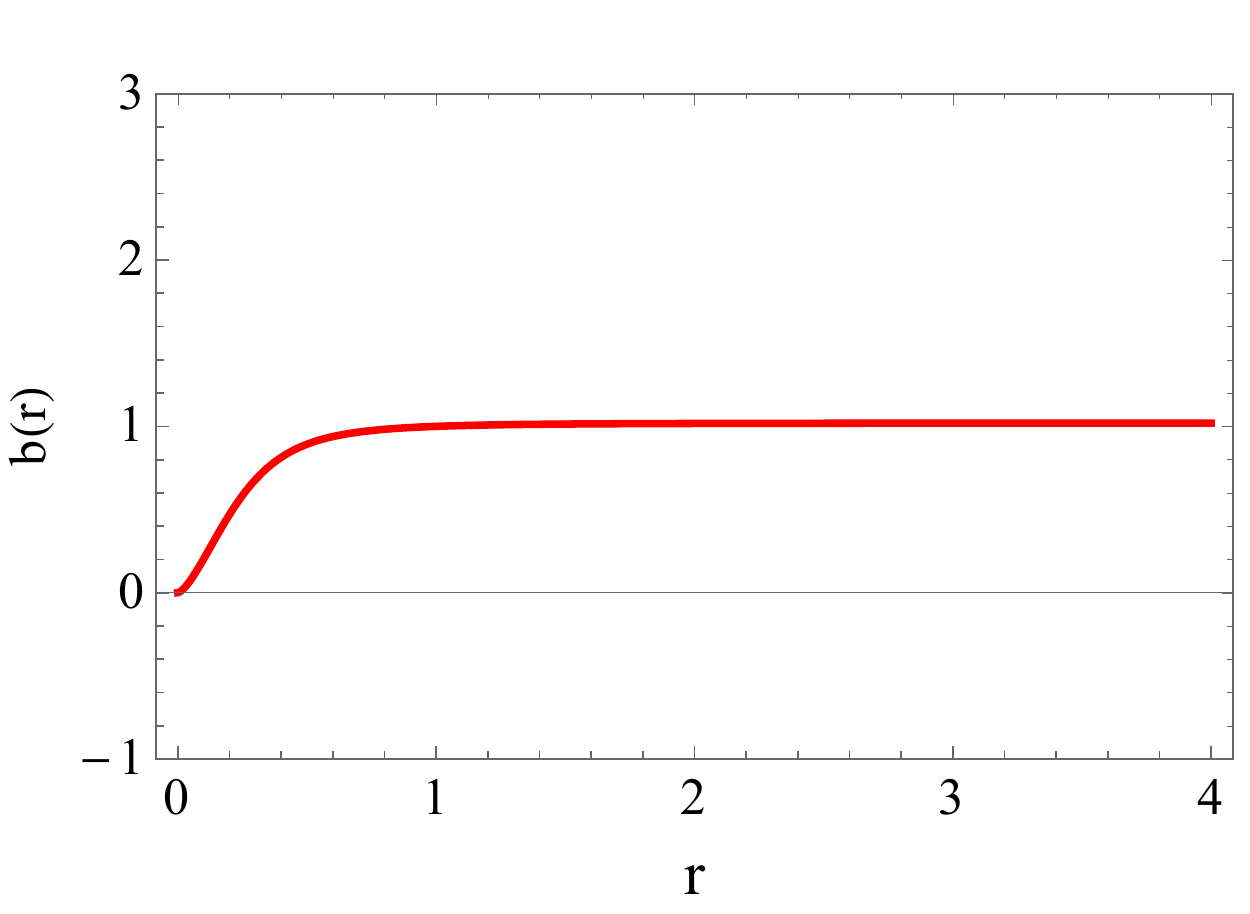}
\includegraphics[width=8.0 cm]{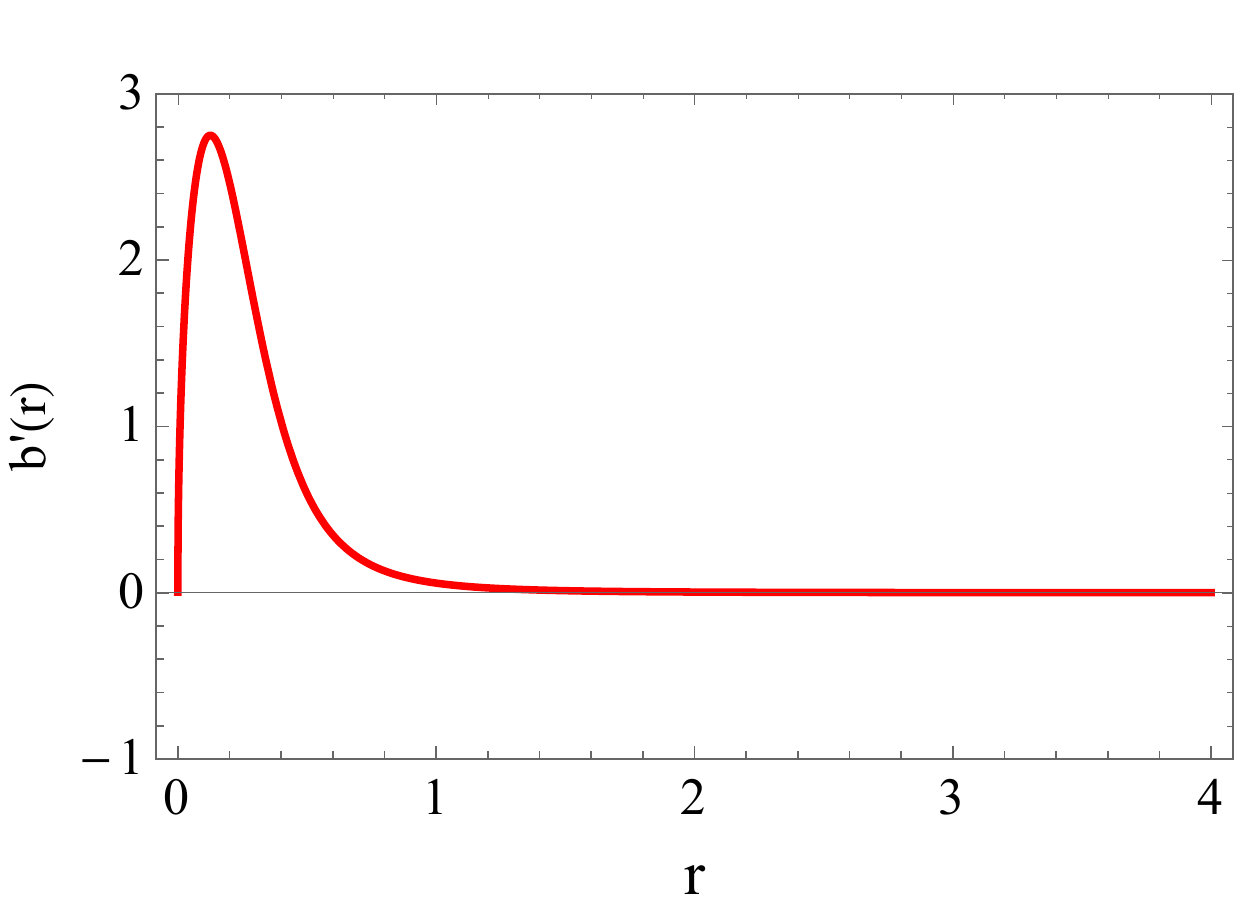}
\caption{Plots for $b(r)$ and $b'(r)$ as a function of $r$ for our exact isotropic wormhole.
The constants are $\alpha=0.1$, $\omega=0.5$ and $r_0=1$ for the left and right plots, respectively.}\label{f1}
\end{figure*}

\begin{figure*}
\includegraphics[width=8.0 cm]{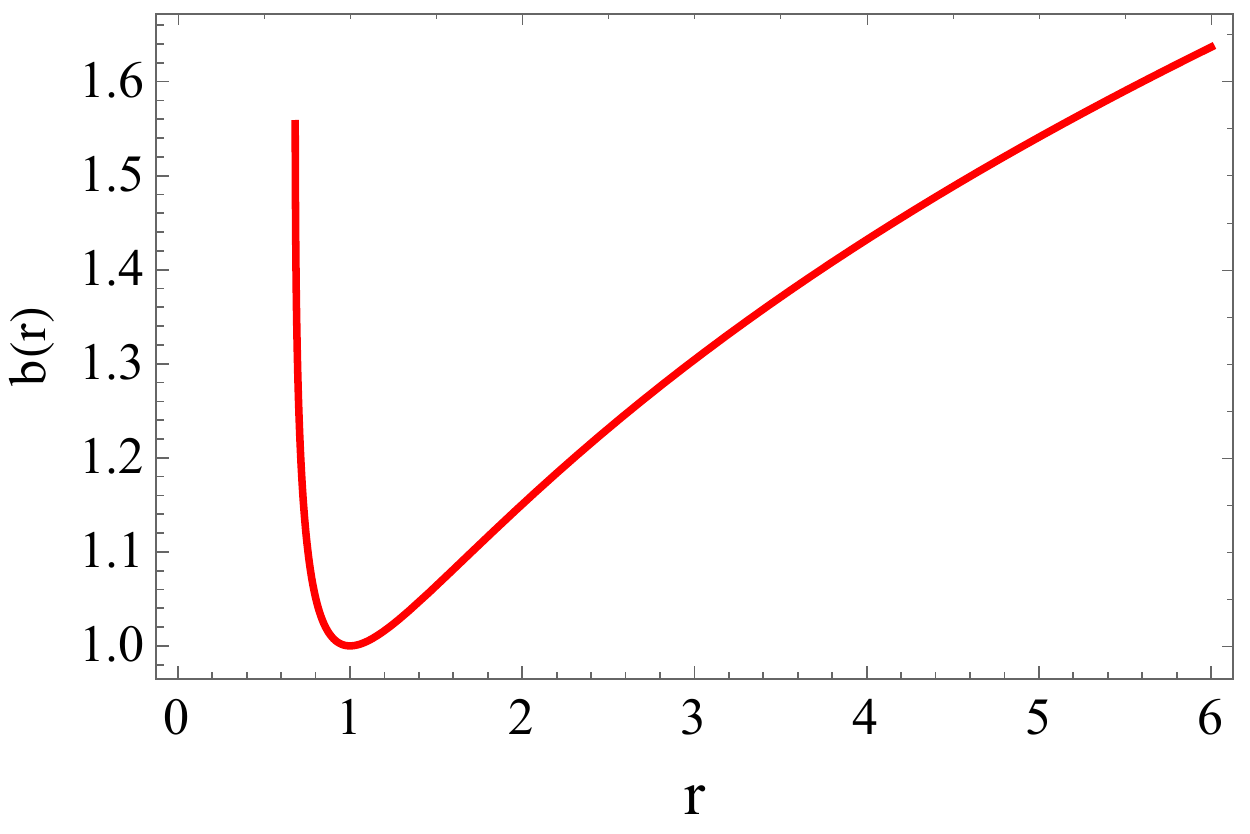}
\includegraphics[width=8.0 cm]{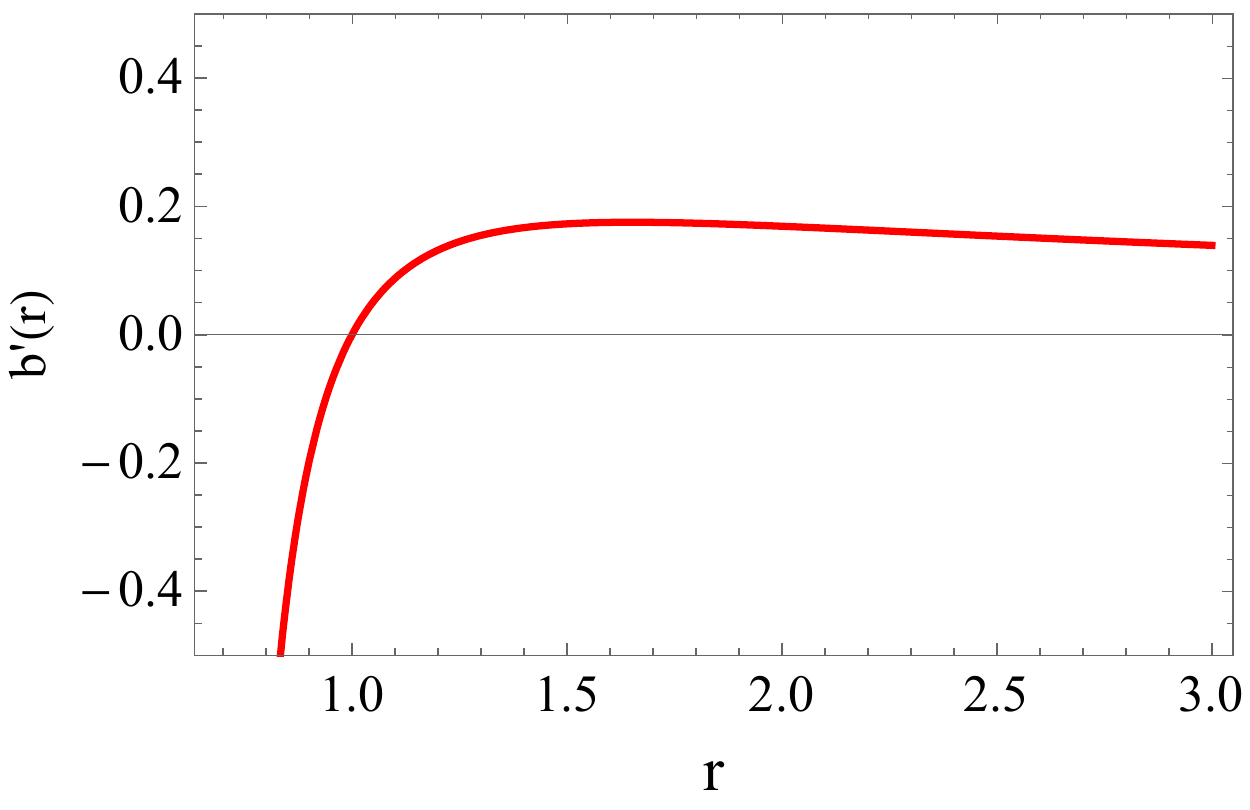}
\caption{Representation of $b(r)$ and $b'(r)$ as a function of $r$ for the anisotropic wormhole. 
The constants are $\alpha=0.1$ and $r_0=1$ along with $\omega_t=-1/3$ for left and right plots, respectively.}\label{f2}
\end{figure*}

\begin{figure*}
\includegraphics[width=8.0 cm]{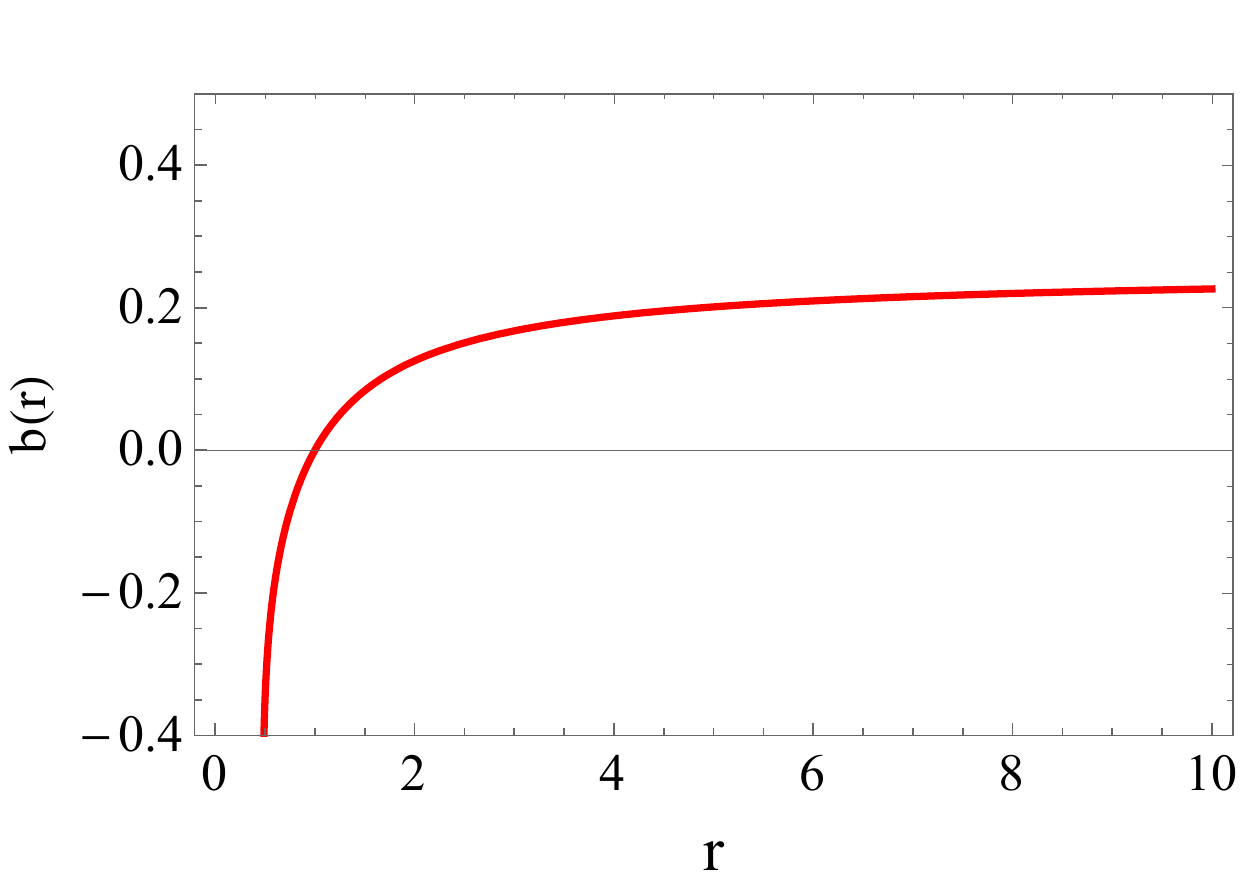}
\includegraphics[width=8.0 cm]{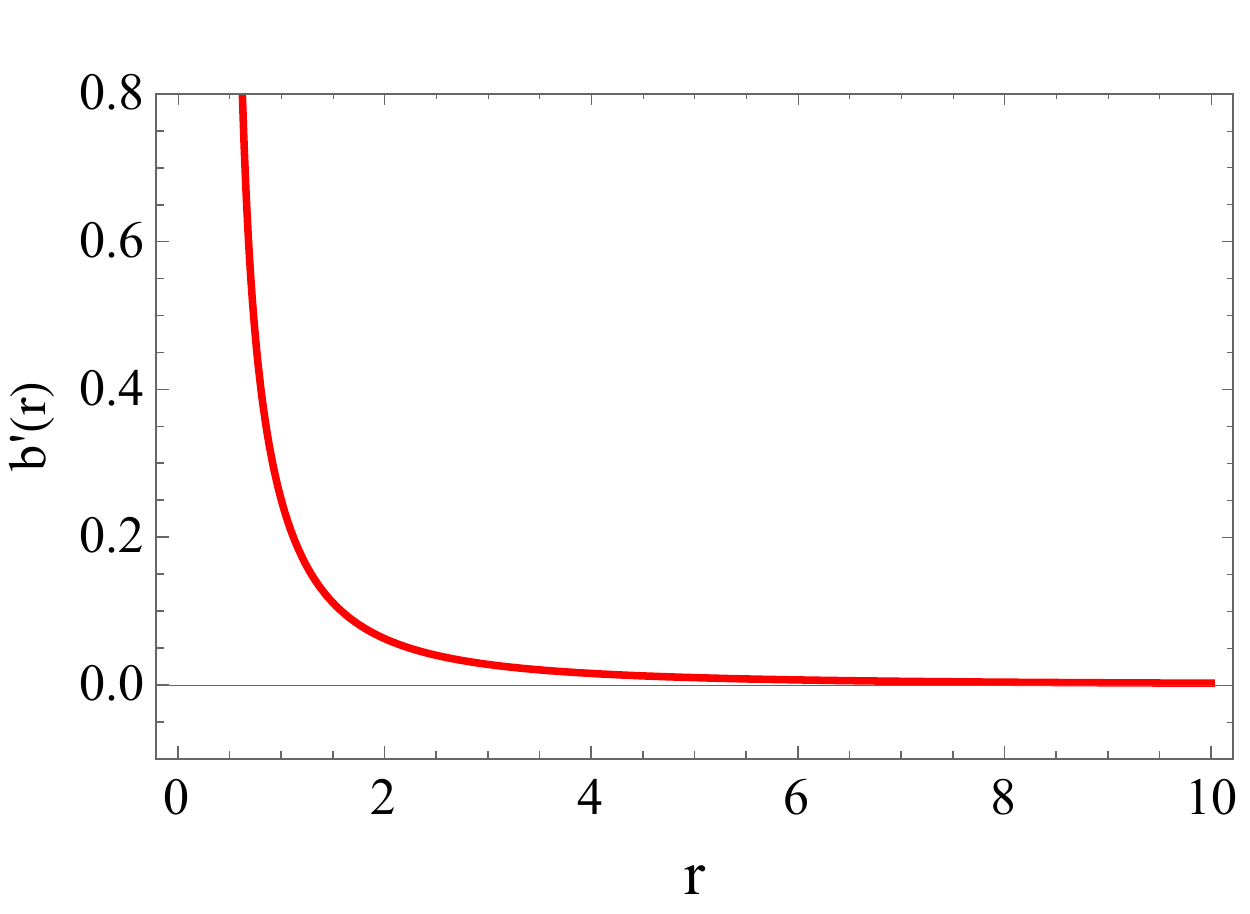}
\caption{Plots for $b(r)$ and $b'(r)$ as a function of $r$ for our assumed model $\rho(r)=\rho_0 \left(  \frac{r_0}{r} \right)^{\beta}$.
The constants are $\alpha=0.1$, $\rho_0=0.01$ and $r_0=1$ along with $\beta=4$ for the left and right plots, respectively.}\label{f3}
\end{figure*}

\subsection{Model with \texorpdfstring{$\rho(r)=\rho_0 \left(  \frac{r_0}{r} \right)^{\beta}$}{Lg}}
One can also consider a specific energy density to obtain the shape function. Here, we shall consider the following energy density profile given in \cite{Lobo:2012qq}:
\begin{equation}
    \rho(r)=\rho_0 \left(  \frac{r_0}{r} \right)^{\beta},
\end{equation}
where $\beta$ and $\rho_0$ are constants. Solving the Eq. (\ref{DRE1}), we obtain the following shape function
\begin{equation}\label{eq24}
    b(r)=-\frac{r^3}{\alpha}\left(1 \pm \sqrt{1+ \frac{4 \alpha \left(C(3-\beta)+ 8 \pi r^3 \rho_0 ( \frac{r_0}{r})^{\beta}   \right) }{r^3}}   \right),
\end{equation}
in which $C$ is a constant of integration. Using the condition $b(r=r_0)=r_0$, we obtain
\begin{equation}
    C=\frac{8 \pi \rho_0 r_0^3}{\beta-3}.
\end{equation}
Substituting the above expression into the Eq. (\ref{eq24}), one obtains the two different branches of the solution, which are
\begin{equation}
     b(r)=-\frac{r^3}{\alpha}\left(1 \pm \sqrt{1+ \frac{32 \pi \rho_0 \alpha  \left(-r_0^3+ r^3 ( \frac{r_0}{r})^{\beta}   \right) }{r^3}}   \right).
\end{equation}
Note that the branch with $-$ve sign is stable and we chose as the physical solution, provided $\beta \geq 4$. On the other hand $\beta=3$, there is an apparent singularity. Notice that there is a third case when $\beta=1,2$, where the sign in the solution flips. The $-$ve  branch solutions for $\beta \geq 4$ are asymptotically flat, for example  by choosing $\beta=4$ one can see this by considering the limit $\alpha \to 0$, the  $-$ve  gives
\begin{equation}
    \frac{b(r)}{r}=\frac{8 \pi \rho_0 r_0^3 (r-r_0)}{r^2}+ \ldots
\end{equation}
 On the other hand, in the limit $\alpha \to 0$, the  $+$ve  goes over
\begin{equation}
    \frac{b(r)}{r}=-\frac{r^2}{\alpha}-\frac{8 \pi \rho_0 r_0^3 (r-r_0)}{r^2}+ \ldots,
\end{equation}
which corresponds to the wormhole solution in a de-Sitter/ anti-de Sitter spacetimes. Finally one can check that for $\beta=1,2$ our solutions are not asymptotically flat. Using the shape function (\ref{eq24}) the wormhole metric defines by the expression
\begin{equation}
    ds^2=-dt^2+\frac{dr^2}{1+\frac{r^2}{\alpha}\left(1- \sqrt{1+\zeta}  \right)}+r^2d\Omega^2_2,
\end{equation}
where
\begin{equation}
    \zeta=\frac{32 \pi \rho_0 \alpha  \left(-r_0^3+ r^3 ( \frac{r_0}{r})^{\beta}   \right) }{r^3}, \notag
\end{equation}
provided $\beta \geq 4$. As we pointed out one should be careful as we already pointed out when $\beta=1,2$, the sign in the solution flips. As a special case one can consider the limit $\beta \to 0$, in that case the energy density is a constant quantity i.e. $\rho=\rho_0$.  As a special example one can consider a wormhole  supported by vacuum energy, i.e. $\rho_0=\rho_{vac}$, which can be described by the energy-momentum tensor
written as \cite{Lemos:2003jb,Carroll:2000fy}
\begin{equation}
    T^{(vac)}_{\mu \nu}=-g_{\mu \nu} \rho_{vac}.
\end{equation}
In fact it quite easy to observe that the effect of an energy-momentum tensor with a cosmological constant is obtained by moving the $\Lambda g_{\mu \nu}$ term to the right-hand side in our field equation (\ref{GBeq}). The vacuum can therefore be thought of as a perfect fluid \cite{Lemos:2003jb,Carroll:2000fy}
\begin{equation}\label{eq13}
    \rho_{vac}=\frac{\Lambda}{8 \pi}
\end{equation}
and
\begin{equation}
    p_{vac}=-\rho_{vac}.
\end{equation}

One can easily get the expression for shape function, inserting the Eq. (\ref{eq13}) to Eq.~(\ref{DRE1}), and the $b(r)$ is
\begin{equation}\label{solb}
    b(r)=-\frac{r^3}{2 \alpha}\left( 1 \pm \sqrt{1+\frac{4 \Lambda \alpha}{3 r}-\frac{4\alpha (\Lambda r_0^4-3 r_0^2-3 \alpha)}{3 r^3 r_0}}  \right).
\end{equation}
To this end, we need a specific EoS such that the Eq. (\ref{n12}) will be satisfied. Note that due to the presence of the cosmological constant our solution (\ref{solb}) is not asymptotically flat. Same situation has precisely pointed out in our articles \cite{casimir1,casimir2}, where solutions are supported by quantum effects using the Casimir energy and GUP corrected Casimir energy.

\begin{figure*}
\includegraphics[width=8.0 cm]{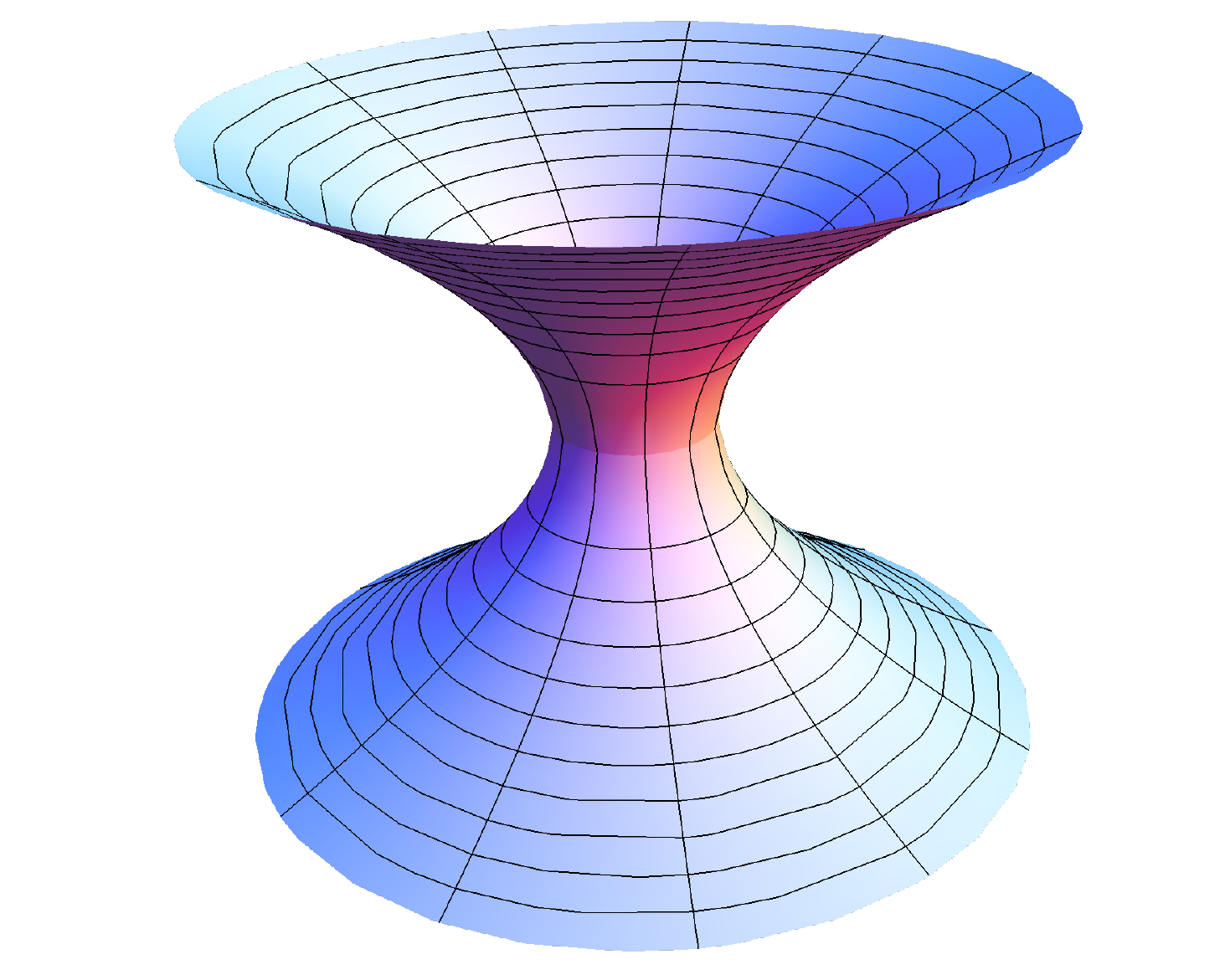}
\includegraphics[width=8.0 cm]{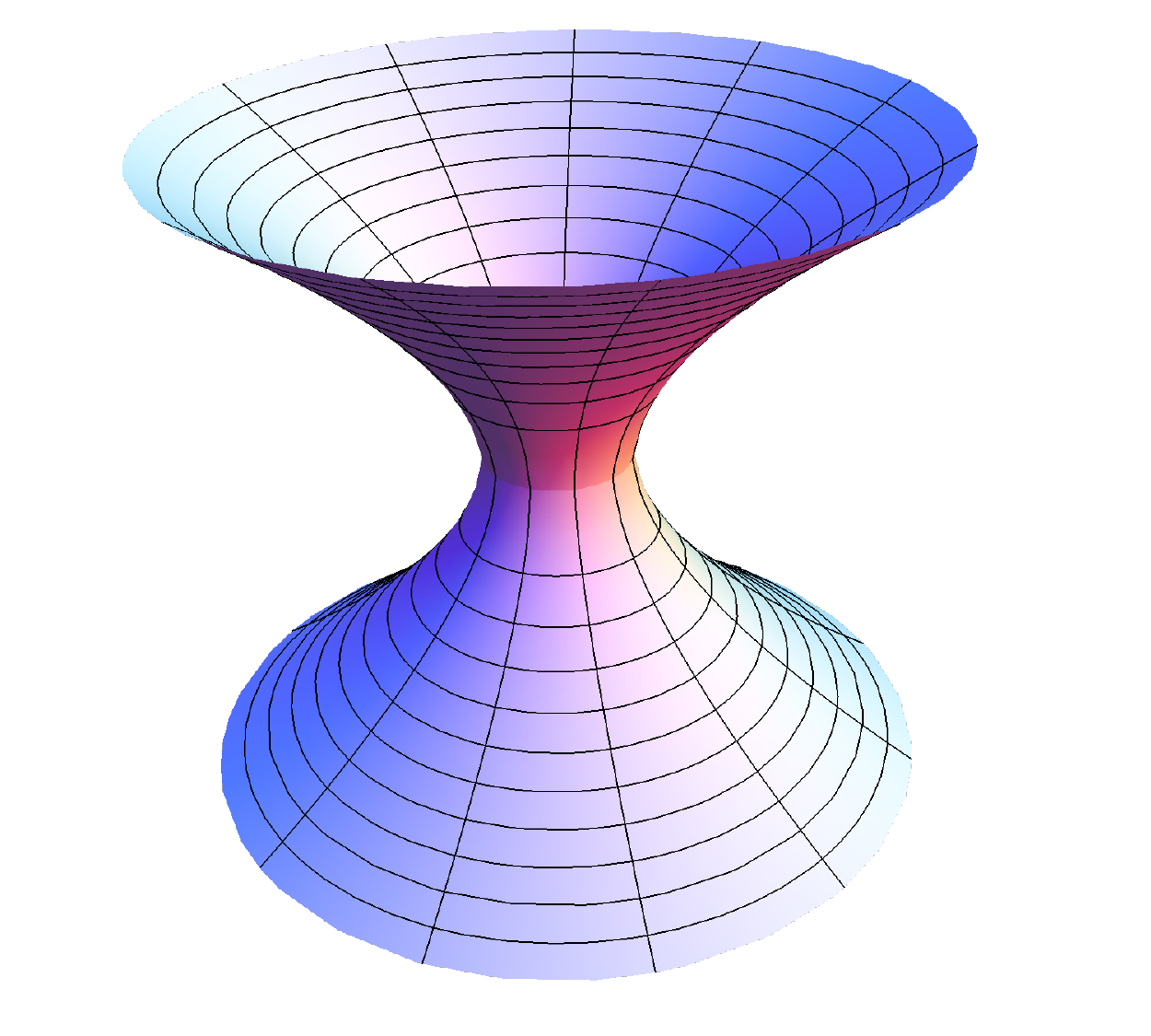}
\caption{The embedding diagram of wormholes geometry along the equatorial plane ($t$ = const, $\theta=\pi/2$). The specific case of a constant redshift $\Phi'(r)=0$  with $\alpha=0.1$ and $r_0=1$, we have drawn embedding diagram for isotropic i.e model A and anisotropic wormhole i.e. model B  in the left and right diagram, respectively. We have considered the 
numerical values of $\omega=0.5$ and  $\omega_t= -1/3$.}\label{f4}
\end{figure*}

\subsection{Model with \texorpdfstring{$b(r)=r \left(  \frac{r_0}{r} \right)^{n+1}$}{Lg}}
One can simplify the mathematical formalism by
assuming a well known shape function with the following form
\begin{equation}
    b(r)=r \left(  \frac{r_0}{r} \right)^{n+1},
\end{equation}
where $n$ is a constant such as: $n=0, 1, 1/2. -1/2, ...$. Next, substituting $b(r)$ in Eq.~(\ref{DRE1})  and solving for $\rho$, we obtain 
\begin{equation}
   \rho(r)=-\frac{r_0 \left[ 2 r_0 \alpha (n+\frac{3}{2})(\frac{r_0}{r})^{2n}+n r^3 (\frac{r_0}{r})^{n}  \right]}{8 \pi r^6}.
\end{equation}
With these setups, the radial and transverse pressures in  (\ref{DRE2}) and (\ref{DRE3}) are leads to
\begin{equation}
    \mathcal{P}_r=- \frac{r_0 \left( (\frac{r_0}{r})^2 r^3-(\frac{r_0}{r})^{2n} \alpha r_0   \right)}{8 \pi r^6},
\end{equation}
and
\begin{equation}
    \mathcal{P}_t=-\frac{(\frac{r_0}{r})^{n+1} \left[ 2 \alpha (\frac{r_0}{r})^{n+1} n-r^2 n +6 \alpha (\frac{r_0}{r})^{n+1}-r^2  \right]}{16 \pi r^4},
\end{equation}
respectively. In the present work we are going to focus on $n=1$.  From Figs. (\ref{f1}-\ref{f3}), we observe that the geometrical flare-out condition is satisfied at the wormhole throat, i.e., $b'(r_0)<1$. In all plots the wormhole throat is set to  one, i.e. $r_0=1$. Again in general we need some EoS in order to satisfy the trace of the field equation that we already discussed. 


\section{Wormhole mass function and the effect of $\alpha$ on the wormhole throat}\label{mf}

In order to estimate the effect of the parameter $\alpha$ on the wormhole throat we precede as follows. From  the first field equation (\ref{DRE1}) we can express $b(r)$ as follows 
\begin{equation}\label{equ46}
    b(r)= -\frac{r^3}{2 \alpha}\left(1 \pm \sqrt{1+ \frac{32 \alpha \pi }{r^3}\int_{r_0}^{r} \rho(r') r'^2 dr' +\frac{4 b_0 \alpha}{r^3}}   \right).
\end{equation}
In the last equation we have identified integrating  constant $C$ with the wormhole throat radius, i.e. $C=b_0$. We can chose the $-ve$ branch of solution and  taking the series expansion around $\alpha$, we obtain the Morris-Thorne solution as a special case
\begin{equation}\label{equ47}
    b(r)=b_0+8 \pi \int_{r_0}^{r} \rho(r') r'^2 dr'+...
\end{equation}
Now, the wormhole mass function is related to the shape function as follows $b(r)=2\, m(r)$. Then the total mass of the wormhole is given by:
\begin{equation}
   M=\lim_{r\to \infty} m(r).
\end{equation}
For the Morris-Thorne wormhole the mass function we obtain
\begin{equation}
    m^{MTH}(r)=\frac{b_0}{2}+4 \pi \int_{r_0}^{r} \rho(r') r'^2 dr'.
\end{equation}
On the other hand, the mass function for the 4$D$ EGB wormhole, one gets 
\begin{equation}
    m(r)=-\frac{r^3}{4 \alpha}\left(1 - \sqrt{1+ \frac{32 \alpha \pi }{r^3}\int_{r_0}^{r} \rho(r') r'^2 dr' +\frac{4 b_0 \,\alpha}{r^3}}   \right),
\end{equation}
Finally, if we express the $b_0$ from Eq. (\ref{equ46}), we obtain 
\begin{equation}
b_0=\underbrace{b(r)-8 \pi \int_{r_0}^{r} \rho(r') r'^2 dr'}_{b^{MTH}_0}+\frac{\alpha b^2(r)}{r^3},
\end{equation}
Using Eq. (\ref{equ47}) we see that 
\begin{equation}
b_0=b^{MTH}_0+\frac{\alpha b^2(r)}{r^3}.
\end{equation}
Hence, at the wormhole throat $b(r_0)=r_0$, we obtain 
\begin{equation}
   b_0=b^{MTH}_0+\frac{\alpha}{b_0}.
\end{equation}
Finally, solve for $b_0$, we have
\begin{equation}
    b_0=\frac{b^{MTH}_0}{2}\pm \frac{\sqrt{(b^{MTH}_0)^2+4 \alpha}}{2}.
\end{equation}
We can accept as a physical solution the positive one. This equation shows that the parameter $\alpha$ increases the wormhole throat compared to the Morris-Thorne case. However, for any $\alpha>0$ we see that the mass function of the wormhole in 4D EGB theory is different from the Morris-Throne wormhole. In other words, having the same mass, means no effect of $\alpha$ on the wormhole throat. We can see that taking the limit $\alpha \to 0$, we have $b_0=b^{MTH}_0$. Moreover, 
one may investigate the behavior of ADM mass as the wormhole geometry requires 
asymptotic flatness spacetime.  Thus, (\ref{metric}) the expression in a more convenient form (4$D$ case), as
\begin{equation}
ds_\Sigma^2=\phi(r) dr^2+r^2\chi(r)
\left(d\theta^2+\sin^2\theta d\varphi^2\right),
\end{equation}
where $\phi(r)= \left(1-\frac{b(r)}{r} \right)^{-1}$ and $K(r)=\chi(r)$. If we follow the approach in \cite{Shaikh:2018kfv}, for the ADM mass reduces to
\begin{equation}\label{a4}
m_{ADM}=\lim_{r\to \infty}\frac{1}{2}\left[-r^2\chi'+r(\phi-\chi)\right].
\end{equation}

Here,  $\chi (r)=1$, which means that for minimum value of the $r$-coordinate $m_{\text{ADM}}=\frac{r_0}{2}$, but this situation is strictly depending on the values of $b(r)$. Here, we
would like to point out that different values of $b(r)$ correspond different ADM mass for the wormholes. Interestingly the two definitions of the wormhole mass, ADM mass and the total mass are in agreement if the corresponding spacetime is asymptotically flat. As our wormhole solutions is not asymptotically flat at $r \to \infty$, but we can consider a flat space in the asymptotic limit by taking into account thin shells.  For instance,  one can cut the original spacetime at a given hypersurface and paste it together with an exterior spacetime leading to a boundary between the two regions with a surface stress-energy tensor.
This procedure must be done imposing the Israel junction conditions \cite{Israel} at the boundary surface $r = R > r_0$,  using the cut-and-paste technique \cite{Poisson:1995sv} to avoid the presence of horizons and singularities.\\
Since, due to the spherical
symmetry the components $g_{\theta\theta}$ and $g_{\phi\phi}$ are already continuous, and so
one is left with imposing the continuity of $g_{tt}$ and $g_{rr}$, by $g_{tt(\text{int})}=g_{tt(\text{ext})}$
and $g_{rr(\text{int})}=g_{rr(\text{ext})}$ at $r = R$. The ranges of $t$ and $r$ are $-\infty <t <+\infty$ and $r_0 \leq r <+\infty$,
where $r_0$ is the radius of the wormhole throat. Here, the interior wormhole solution
is being matched with exterior Schwarzschild geometry. Now, using $1-b(R)/R= 1-2M/R$, we obtain
\begin{equation}
M= \frac{b(R)}{2},
\end{equation}
which is the total mass of the wormhole depending on the values of $b(r)$.
Furthermore, as wormhole is a tunnel-like structure through space and is itself empty of all space and mass. The space around the hole must have some mass, to contain the hole of the wormhole. So, there is no critical mass for the wormholes. However, one may consider the scenario if a wormhole accretes matter,
in that case, there might be a maximal mass of a wormhole accreting
matter and eventually a point of a
'steady state' where the accreted matter all ends up on the other side. This is an interesting point for investigation in the near future and has added further comments. 


\begin{figure*}
\includegraphics[width=8.0 cm]{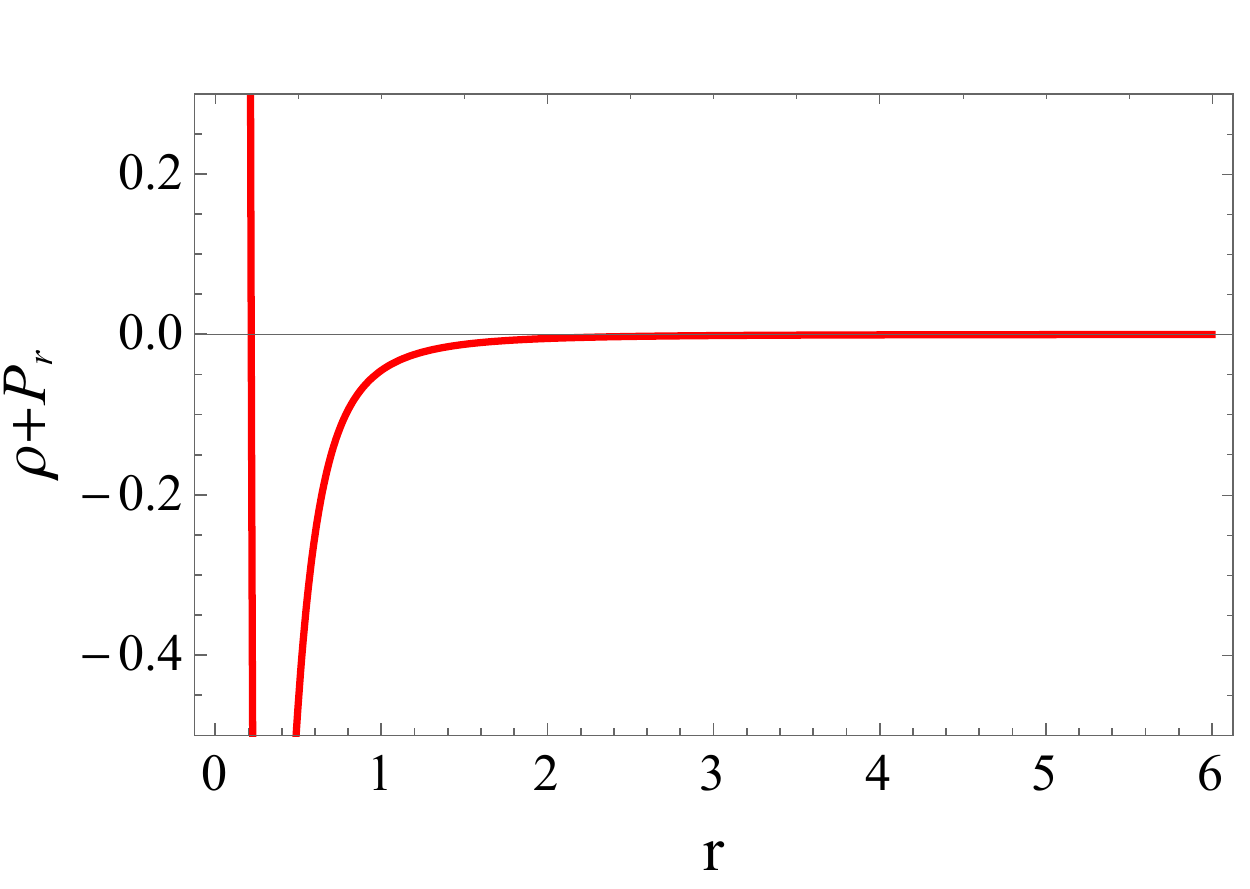}
\includegraphics[width=8.0 cm]{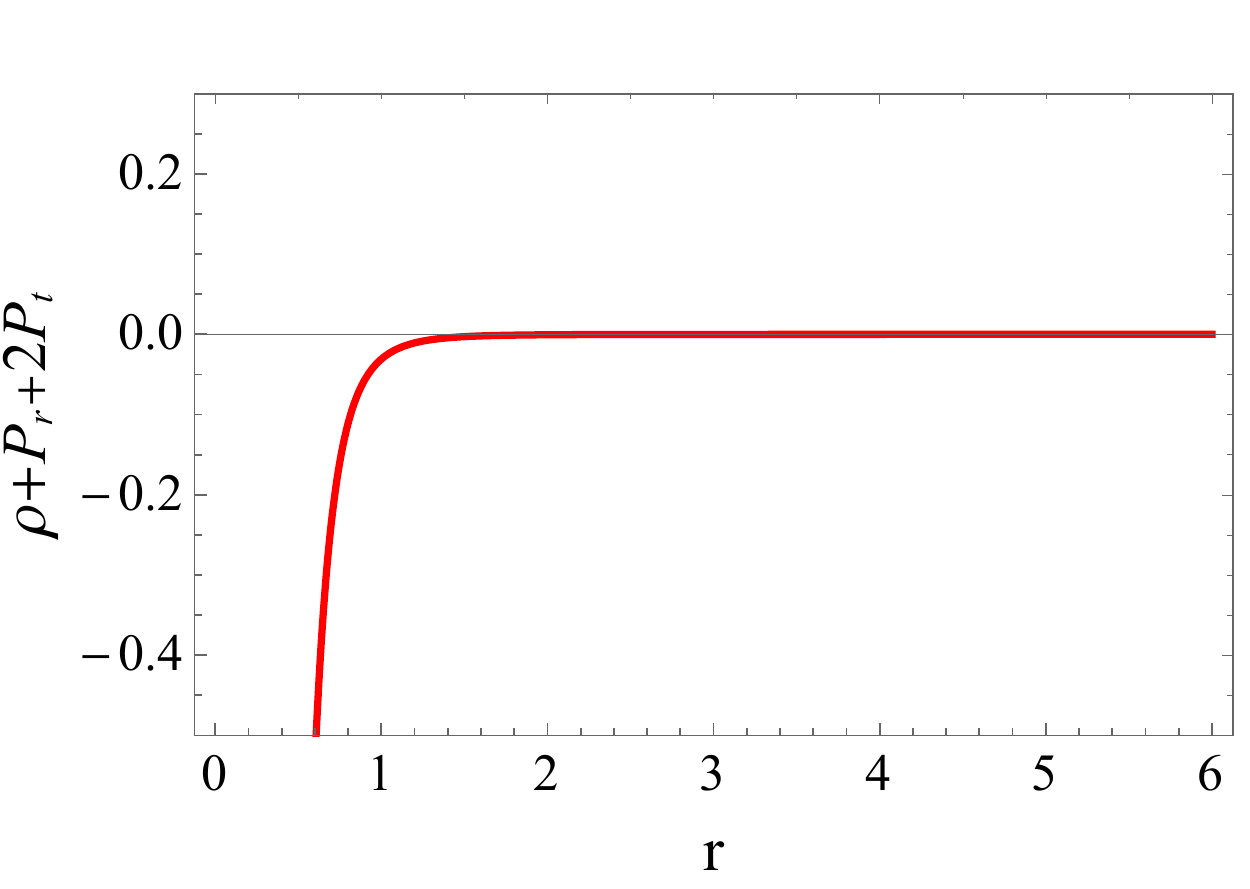}
\caption{The behavior of NEC  and SEC diagrams have been plotted for the isotropic wormhole against $r$.
Here also NEC is violated, this implies WEC is also violated. For plotting the constants are
$\alpha=0.1$ and $r_0=1$ along with $\omega_t=0.5$. 
}\label{f5}
\end{figure*}

 \begin{figure*}
\includegraphics[width=8.0 cm]{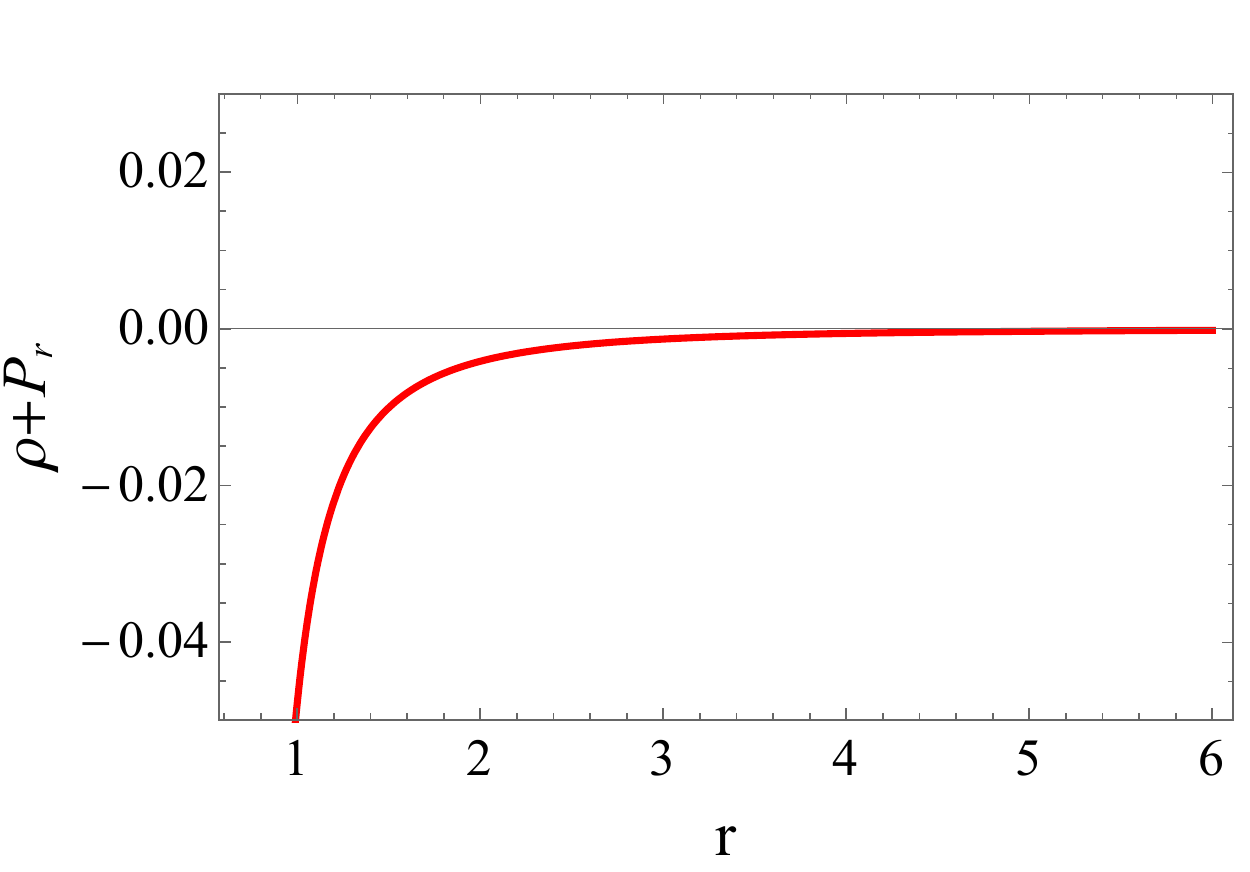}
\includegraphics[width=8.0 cm]{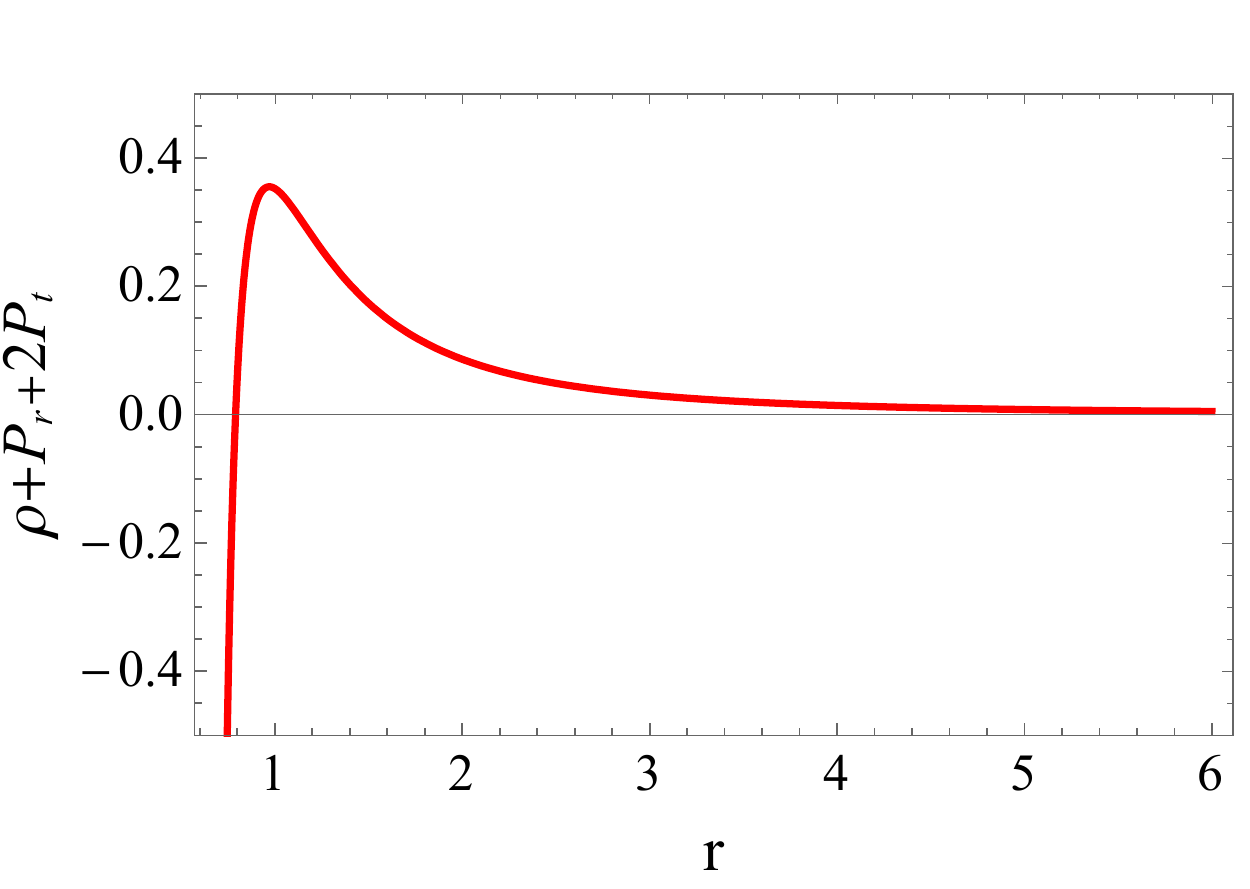}
\caption{The behavior of NEC  and SEC diagrams have been plotted for the anisotropic wormhole against $r$.
Here also NEC is violated, this implies WEC is also violated. For plotting the constants are
$\alpha=0.1$ and $r_0=1$ along with $\omega_t=-1/3$. 
}\label{f6}
\end{figure*}

\begin{figure*}
\includegraphics[width=8.0 cm]{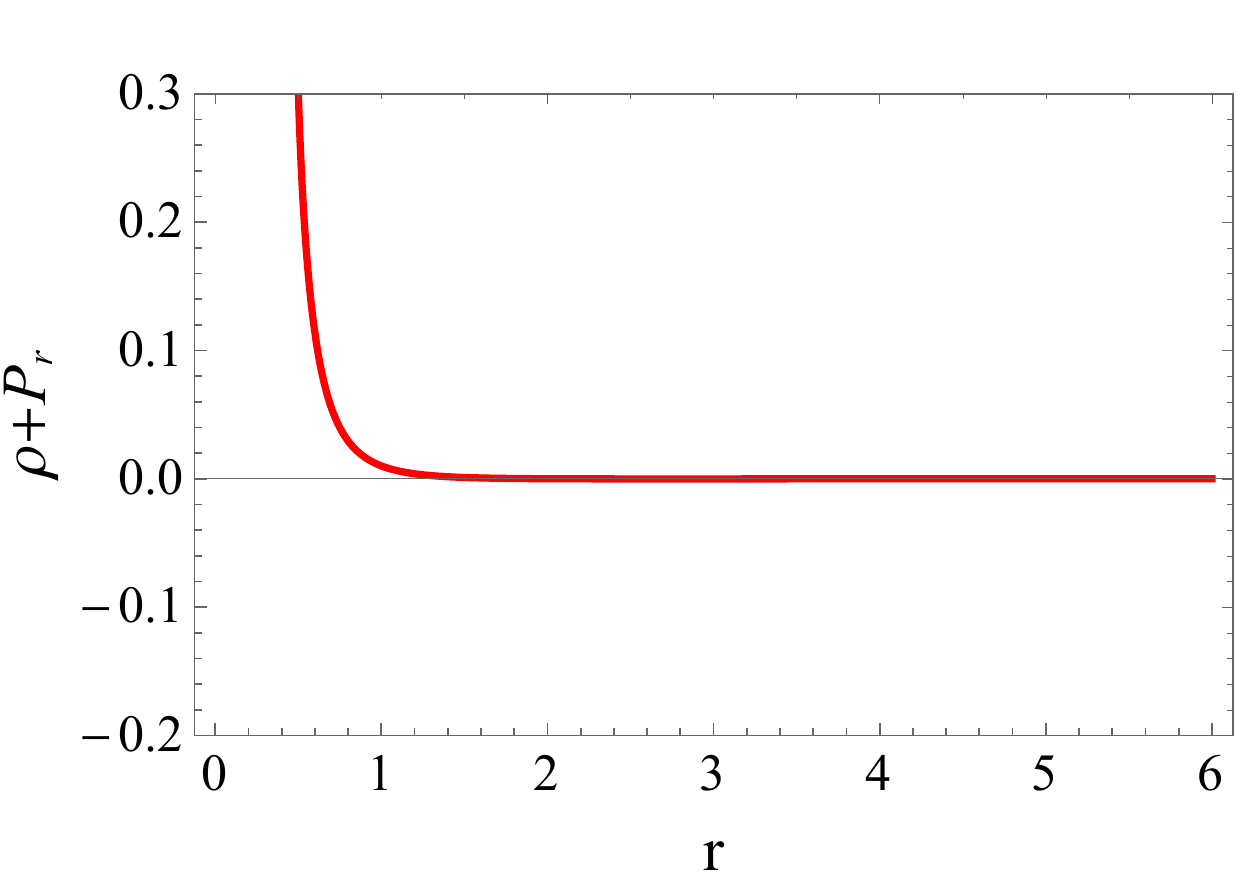}
\includegraphics[width=8.0 cm]{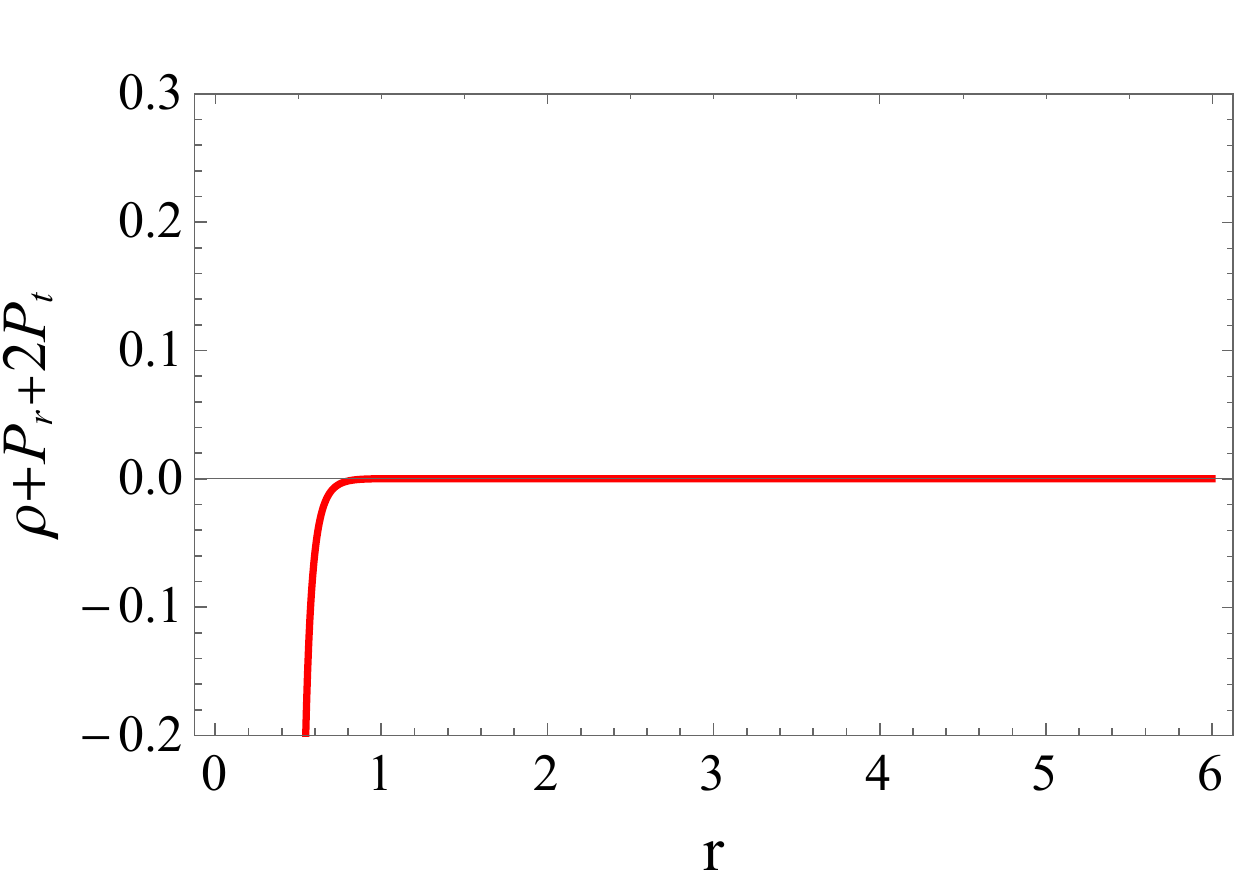}
\caption{Plots for NEC (left panel) and SEC diagrams (right panel) against radial coordinate and depending on
$\rho(r)=\rho_0(r_0/r)^{\beta}$.  In this cases, NEC is violated outside the throat whereas SEC is violated
throughout the spacetime.   We use the numerical values $\alpha=0.1$ and $r_0=1$, $\rho_0=0.01$ along with $\beta=4$, respectively. \label{f7}
}
\end{figure*}

\begin{figure*}
\includegraphics[width=8.0 cm]{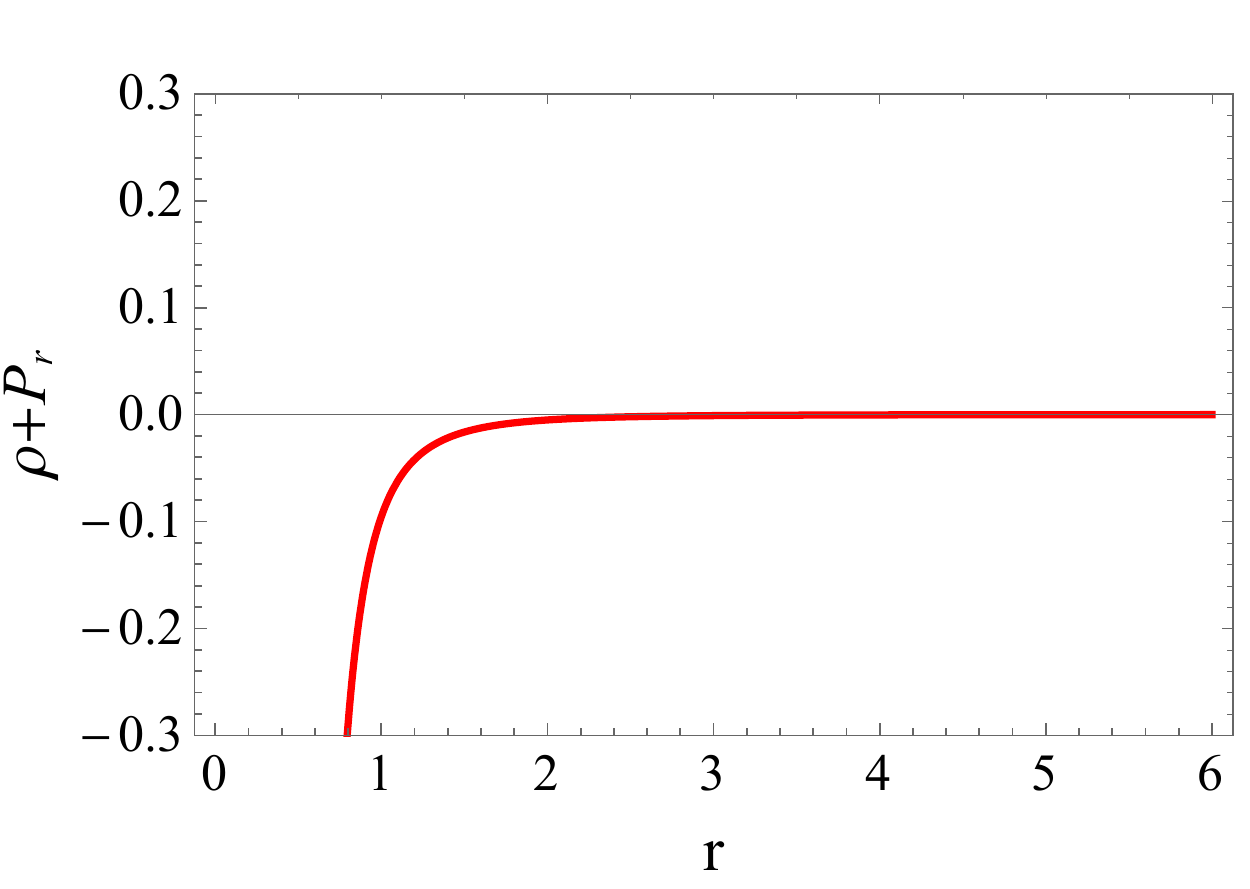}
\includegraphics[width=8.0 cm]{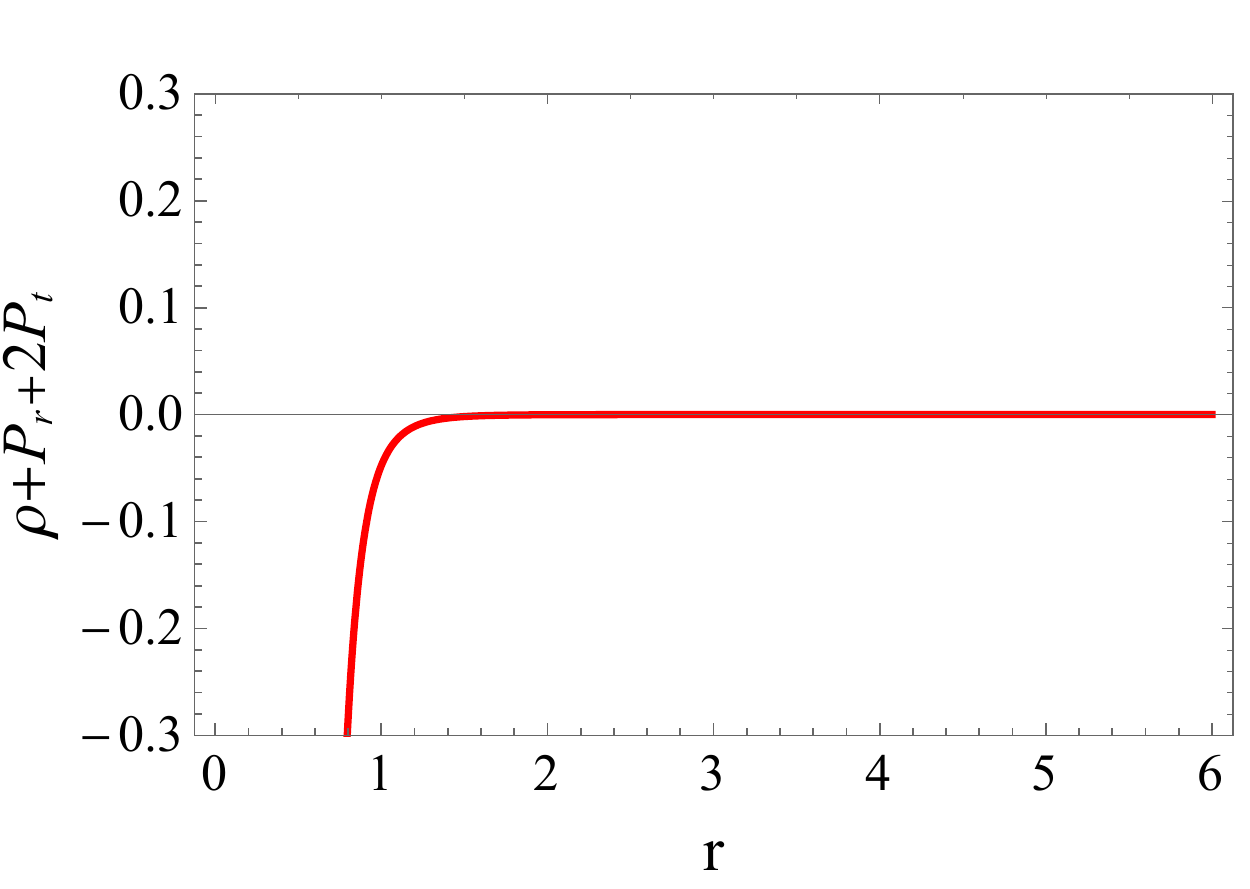}
\caption{Plots for NEC and SEC diagrams in the left and right panel against radial coordinate depending on
the shape function  $b(r)=r(r_0/r)^{n+1}$. The results seem NEC and SEC are violated throughout the spacetime. We use the numerical values 
 $\alpha=0.1$ and $r_0=1$ along with $n=1$, respectively.
}\label{f8}
\end{figure*}
\section{Embedding diagram}\label{sec4}

In this section we will analyze the embedding diagram that helps us to impose the demand of 
the spacetime metric (\ref{metric}) describe a wormhole. Of particular interests the geometry, we 
consider an equatorial slice $\theta=\pi/2$ at some fix moment in time $t=const$. 
With this constraint the metric (\ref{metric}) becomes,
\begin{equation}\label{emb}
    ds^2=\frac{dr^2}{1-\frac{b(r)}{r}}+r^2d\phi^2.
\end{equation}
The reduced metric (\ref{emb}) can be embedded into a $3$-dimensional
Euclidean space, and in cylindrical coordinates $r$, $\phi$ and $z$ has the form
\begin{equation}
    ds^2=dz^2+dr^2+r^2d\phi^2.
\end{equation}
The embedded surface $z(r)$ can be obtained by reversing and integrating 
from the last two equations, we obtain the slope
\begin{equation}
    \frac{dz}{dr}=\pm \sqrt{\frac{r}{r-b(r)}-1}.
\end{equation}

With the illustration of Fig. \ref{f4},  we  explore the geometrical properties of these matrices (\ref{metricw}) and (\ref{eq39}) via the embedding diagram. Numerical values are enlisted in the caption of Fig. \ref{f4}.

\begin{figure*}
\includegraphics[width=8.0 cm]{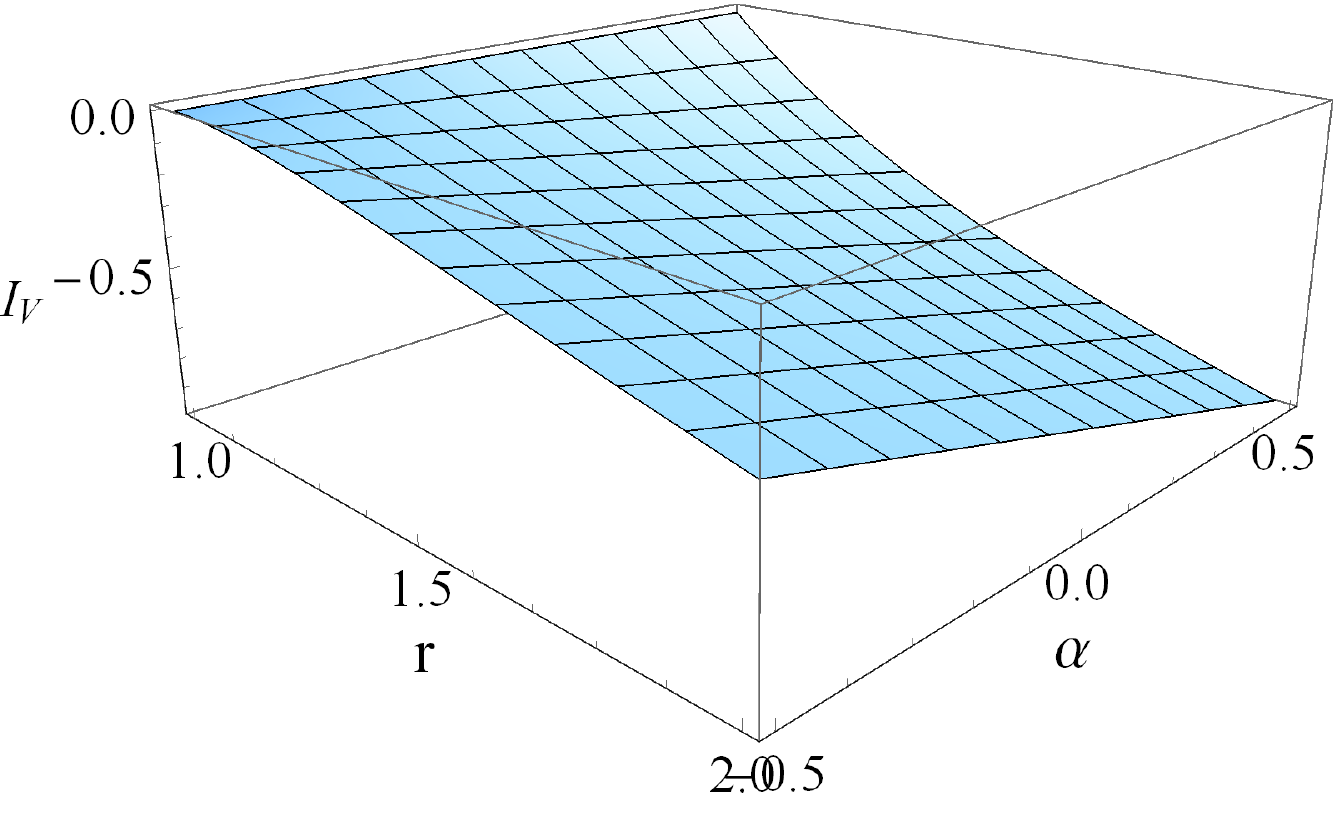}
\includegraphics[width=8.0 cm]{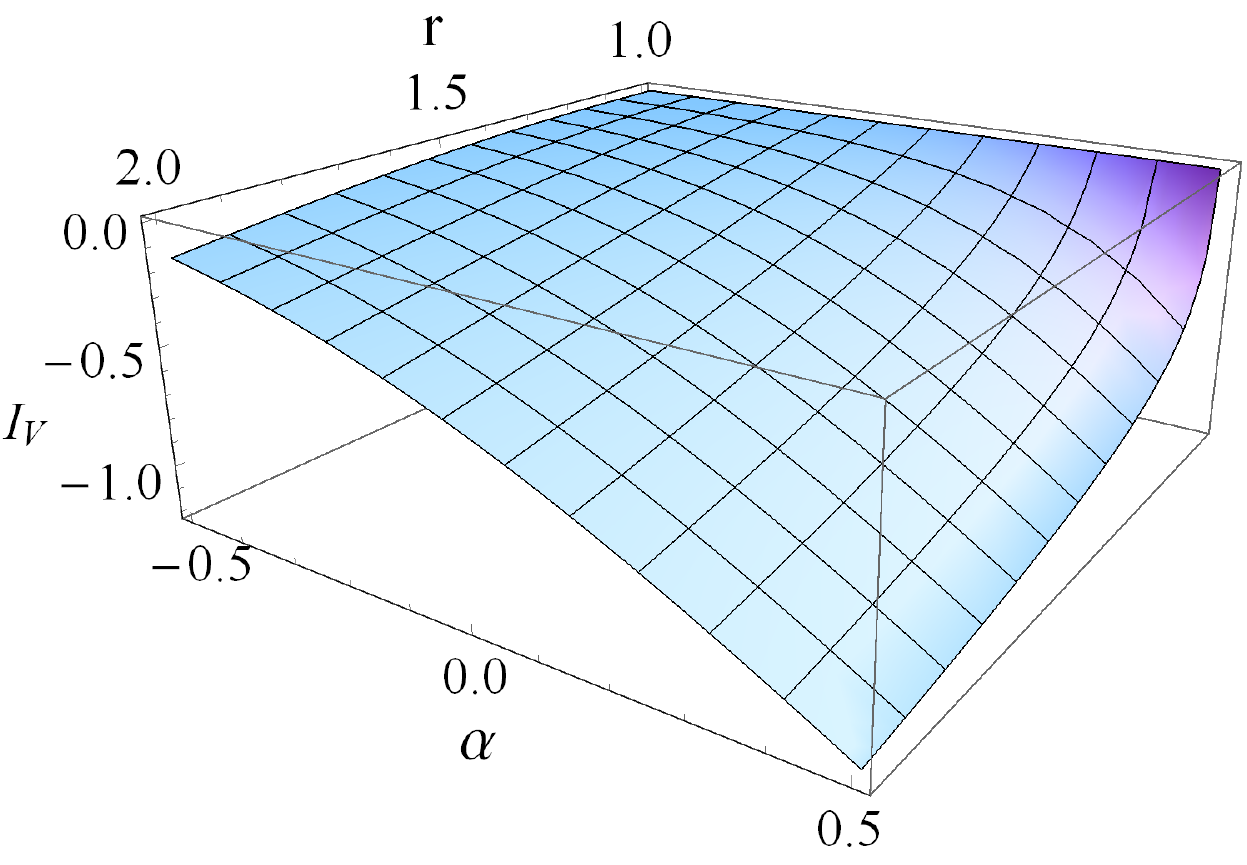}
\caption{3D plots for `volume-integral quantifier' associate with model A \& B in Sec. \ref{sec3}. 
Interesting to find that when $a \to r^{+}_0$ then $\mathcal{I}_V \to 0$, i.e. minimize the
violation of energy conditions would be possible. We consider the same set of values as of Figs. \ref{f1} and \ref{f2} with $a=2$. 
}\label{f9}
\end{figure*}

\begin{figure*}
\includegraphics[width=8.0 cm]{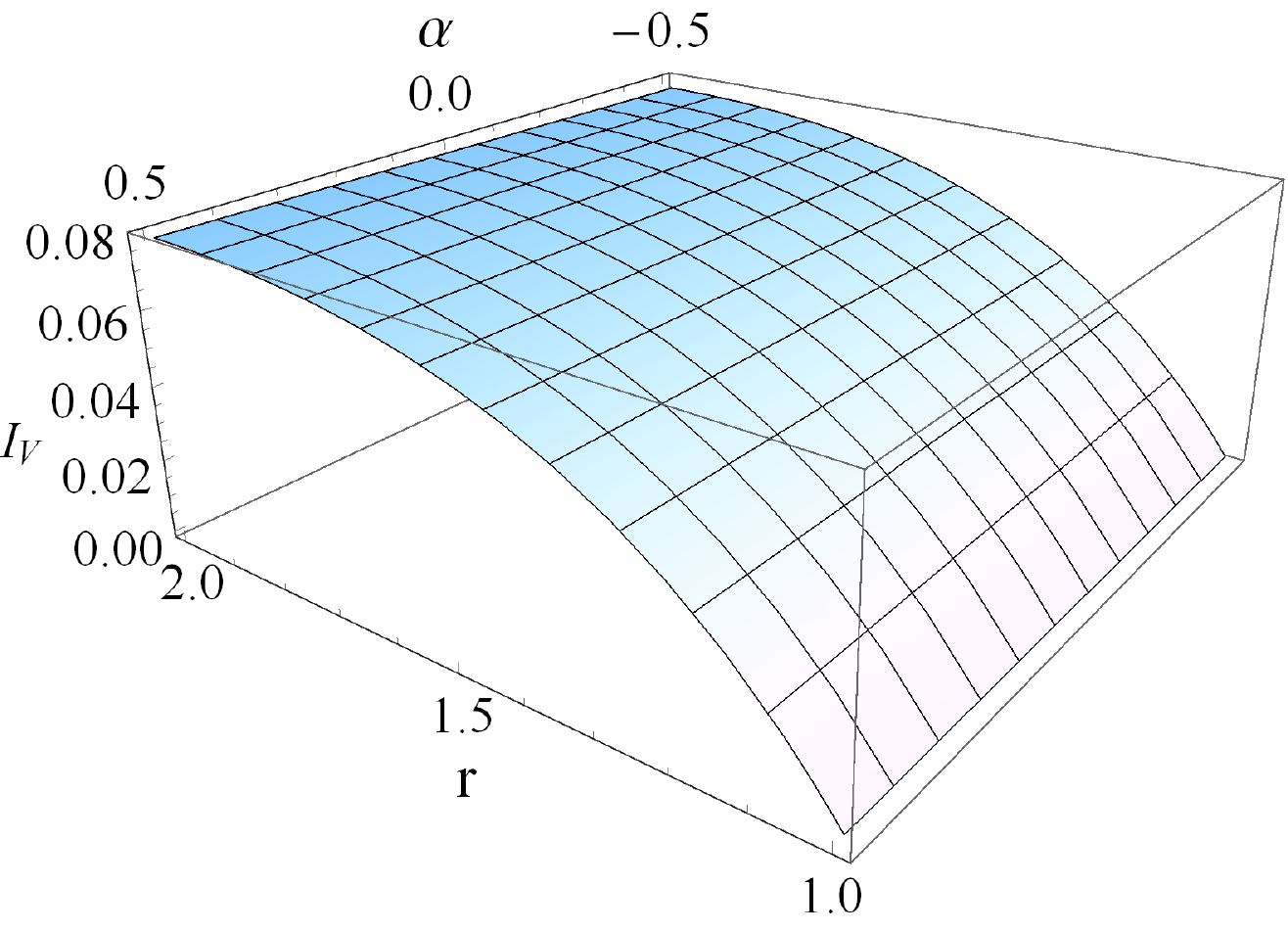}
\includegraphics[width=8.0 cm]{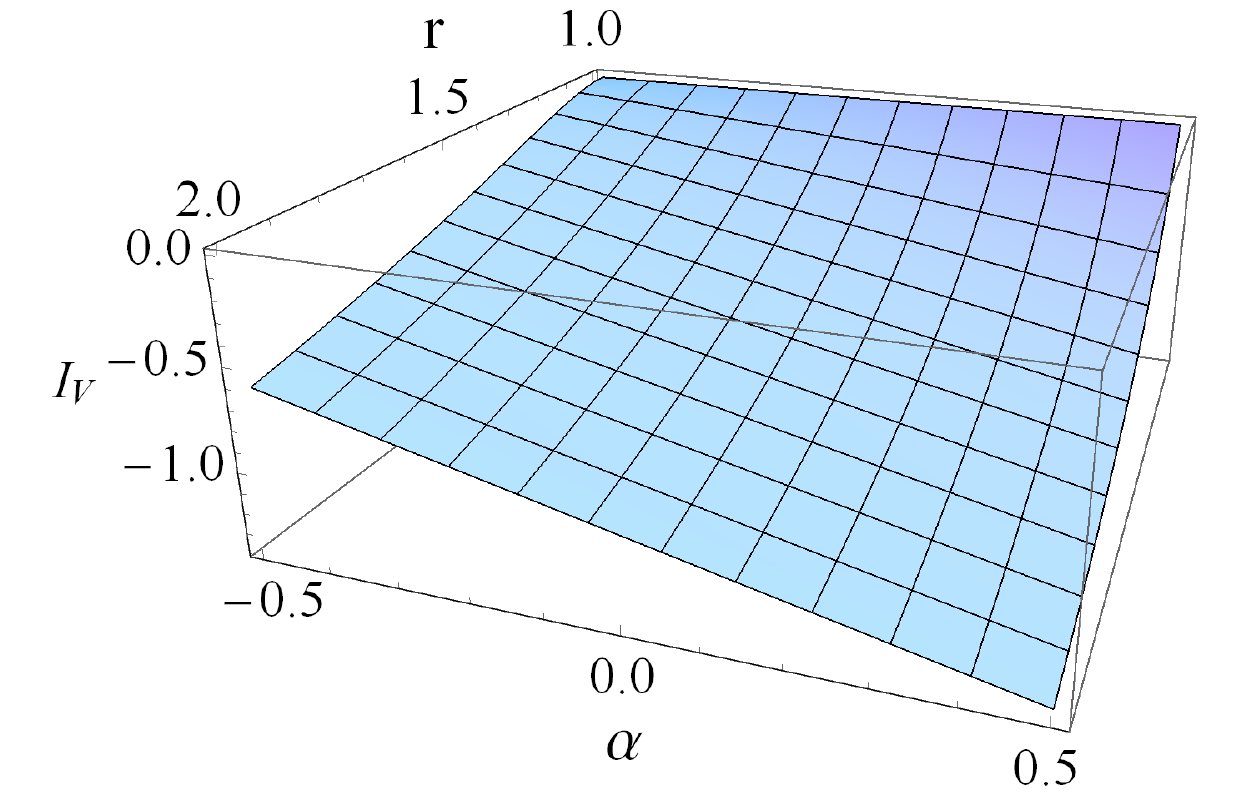}
\caption{
The plot depicts the `volume-integral quantifier' associate with model C \& D in Sec. \ref{sec3}.
It is clear that when $a \to r^{+}_0$ then $\mathcal{I}_V \to 0$, also. We consider the same set of values as of Figs. \ref{f3} and \ref{f4} with $a=2$.}\label{f10}
\end{figure*}

\section{Energy Conditions}\label{sec5}

In this section we present a detail description of the energy conditions, that are  sets of inequalities depending on energy momentum tensor. To be specific, we start by finding 
wormhole solutions for weak energy condition (WEC), i.e. $T_{\mu\nu} U^{\mu}U^{\nu}$, where $U^{\mu}$ is
a timelike vector. For the given diagonal EM tensor, the WEC
implies
\begin{equation}\label{EC1}
\rho (r)\geq 0 ~~\text{and}~~ \rho (r)+\mathcal{P}_{i}(r)\geq 0,
\end{equation}
In next,  the null energy condition (NEC) is given by $T_{\mu\nu} k^{\mu} k^{\nu}$, where $k^{\mu}$ is null vector.
The NEC for diagonal EM tensor implies that
\begin{equation}\label{EC2}
\rho (r)+\mathcal{P}_{i}(r)\geq 0, ~~i= 1,2,3
\end{equation}
wereas strong energy condition (SEC) asserts that $\left(T_{\mu\nu}-\frac{1}{2} T g_{\mu\nu}\right) U^{\mu} U^{\nu}\geq 0$ for any  timelike vector $U^{\mu}$.  The strong energy condition (SEC) asserts that gravity is attractive,
\begin{equation}\label{EC3}
\rho (r)+ \sum \mathcal{P}_{i}(r)\geq 0, ~~\text{and}~~ \rho (r)+\mathcal{P}_{i}(r)\geq 0.
\end{equation}
Note that the WEC or SEC imply NEC, but it follows that any violation of the NEC also violates the SEC, WEC, and DEC. 

Now we consider, the reduced NEC for master Eqs. (\ref{DRE1}-\ref{DRE3}),
when evaluated at the throat is given by 
\begin{equation}
    \rho (r)+\mathcal{P}_{r}(r)|_{r=r_0}=\frac{\left( r^3+2\alpha b(r) \right)(r b'(r)-b(r))}{8 \pi r^6}|_{r=r_0}.
\end{equation}

On the other hand the strong energy condition (SEC) stipulates that
\begin{equation}
\rho (r)+\mathcal{P}_t(r)\geq 0,
\end{equation}
yielding
\begin{equation}
\rho (r)+\mathcal{P}_t(r)|_{r=r_0}=\frac{b(r)\left( r^3-9 \alpha b(r)+4 r \alpha b'(r)  \right)}{8 \pi r^6}|_{r=r_0}.
\end{equation}
Finally using Eq. (\ref{EC3}), we have
\begin{equation}
\rho (r)+\mathcal{P}_{r}(r)+2\mathcal{P}_t(r)\geq 0,
\end{equation}
thereby, at the wormhole throat which yields
\begin{equation}
   \frac{b(r)\left(-4 \alpha b(r)+2 \alpha r b'(r)   \right)}{4 \pi r^6}|_{r=r_0}.
\end{equation}

In Figs. (\ref{f5}-\ref{f8}) we present the energy conditions for all specific cases. In all plots we have chosen a positive value of the Gauss-Bonnet coupling constant $\alpha$, and obtained results suggest that the energy conditions in general are not satisfied at the wormhole throat. Particularly, one can easily see that for $\alpha >0$ the NEC, and consequently the WEC, are violated at the throat, due to the flaring-out condition.

\section{Volume Integral Quantifier}\label{sec6}
The starting point of this discussion is to evaluate the \emph{volume integral quantifier}, which provides information about the `total amount of exotic matter' required for wormhole maintenance. To do this one can
compute the definite integrals
$\int{T_{\mu\nu}k^\mu k^\nu}$ and $\int{T_{\mu\nu}U^\mu U^\nu}$, where $U^\mu $ is the four-velocity \cite{Visser:2003yf,Kar:2004hc}.
The most usual choice is the integral including $\rho$  and $\mathcal{P}_r$, with the following definite integrals $I_V =\int\left(\rho(r)+\mathcal{P}_r(r)\right)\mathrm{d}V$, where $\mathrm{d}V=r^2\sin\theta \mathrm{d}r \mathrm{d}\theta \mathrm{d}\phi$.
In this method, the total amount of exotic
matter is measured by
\begin{eqnarray}
\mathcal{I}_V=\oint [\rho+\mathcal{P}_r]~\mathrm{d}V=2 \int_{r_0}^{\infty} \left(  \rho+\mathcal{P}_r\right)~\mathrm{d}V,
\end{eqnarray}
which can also be written as
\begin{eqnarray}
\mathcal{I}_V =8 \pi \int_{r_0}^{\infty} \left(  \rho+\mathcal{P}_r  \right)r^2  dr.
\end{eqnarray}

Suppose now that the wormhole extends from the throat, $r_0$, with a cutoff of the stress energy tensor at a certain radius $a$, one deduces
\begin{equation}
\mathcal{I}_V=   8 \pi \int_{r_0}^{a} \left(  \rho+\mathcal{P}_r  \right)r^2  dr,
\end{equation}
where $r_0$ is the throat radius and hence the minimum value of $r$. The key point of this discussion is when the limit as $a \to r^{+}_0$, one verifies that $\mathcal{I}_V \to 0$.
For each wormhole solutions we found from Figs. (\ref{f9}-\ref{f10}) that one
may construct wormhole solutions with small quantities of exotic matter, which needs to hold open the wormhole throat. 
This is all in agreement with general theorems on the wormhole energy conditions. According to the  \textit{topological cosmic censorship conjecture} by Friedman, Schleich,
and Witt \cite{Friedman} that any two causal curves extending from past to the future null infinity is homotopy equivalent to each
other. Moreover,  this theorem tells us that in a 
spacetime containing a traversable wormhole the averaged null energy condition must be violated along at least some (not
all) null geodesics, but the theorem provides very limited information on where these violations occur.  Beside of that the essential features of a wormhole geometry are largely encoded in the spacelike section and in the condition for nonexistence of horizons ($g_{00} \neq 0$).  However, our model does violate WEC in some interval of time (but not always). In this sense it is impossible to probe the interior topology actively from far away.\\
\\
Indeed there are counterexamples to passive topological censorship. Now the question is to what extent our wormhole agrees with these notions. If our spacetime is asymptotically anti-de Sitter and not globally hyperbolic the answer is not quite obvious. There are few examples related to cosmic censorship in a Kerr-like phantom wormhole (WH) which contains a singularity that is not protected by an event horizon
\cite{DelAguila:2018gni}. However, these wormholes were not traversable, and furthermore would, in principle, develop some type of singularity also.\\ 
\\
Since, our solution is Lorentzian wormholes possibly through which observers may freely traverse. In this situation, discussion about topological censorship is very different -both mathematically and physically - from the theorem considered in this paper and their discussion is beyond its scope.




\section{Ending Comments}\label{sec7}
In this paper, we explore Morris-Thorne wormholes, i.e., static and spherically symmetric traversable wormholes, in the framework of recently formulated   $4D$ EGB. Throughout our discussion we consider a constant redshift function i.e., $\Phi'(r)= 0$, which simplifies the calculations and provides interesting exact wormhole solutions. Firstly, 
we have found an exact solution in  $4D$ EGB supported isotropic matter source. Importantly we also found a wormhole solution supported by anisotropic matter source with EoS relating two pressure components.  Moreover, we have considered a specific shape function, power law energy density profile. 
Here we have shown that the flare-out condition is satisfied for different models with the positive coupling constant $\alpha$. To this end, we have analyzed the null, weak, and strong conditions at the wormhole throat with a radius $r_0$, and shown that in general the classical energy conditions are violated by some small and arbitrary quantities at the wormhole throat for $\alpha>0$. The GB quadratic curvature terms made a profound influence on the obtained solutions which revealed interesting features. 

Further, the results presented here are a generalization of previous discussions on Moriss-Thorne wormholes of GR which are encompassed as special case in the limit $\alpha \rightarrow 0$. The possibility of  generalization of  wormholes  to rotating case and more general Lovelock gravity theories  \cite{Konoplya:2020qqh}  are interesting problems which are being actively considered. 


\end{document}